\documentclass[12pt,preprint]{aastex}

\def\xhat{{\bf \hat x}}
\def\yhat{{\bf \hat y}}
\def\zhat{{\bf \hat z}}

\def\xvec{{\bf x}}
\def\rvec{{\bf r}}

\def\vvec{{\bf v}}
\def\bvec{{\bf B}}

\def\sgn{{\rm sgn}}

\def\lapprox{\mathrel{\hbox{\rlap{\hbox{\lower4pt\hbox{$\sim$}}}\hbox{$<$}}}}
\def\gapprox{\mathrel{\hbox{\rlap{\hbox{\lower4pt\hbox{$\sim$}}}\hbox{$>$}}}}

\def\tvec{{\bf \hat{t}}}

\newcommand{\be}{\begin{equation}}
\newcommand{\ee}{\end{equation}}

\begin{document}

\title{A Model for Patchy Reconnection in Three Dimensions}

\author{M.G. Linton}
\affil{Naval Research Laboratory, Washington, DC}

\author{D.W. Longcope}
\affil{Department of Physics, Montana State University\\
  Bozeman, Montana}

\begin{abstract}

We show, theoretically and via MHD simulations, how a short 
burst of reconnection localized in three dimensions on a 
one-dimensional current sheet creates a pair of 
reconnected flux tubes. We focus on the post-reconnection
evolution of these flux tubes, studying their velocities
and shapes. We find that slow-mode shocks 
propagate along these reconnected flux tubes, releasing magnetic energy 
as in steady-state Petschek reconnection. The geometry
of these three-dimensional shocks, however, differs 
dramatically from the classical two-dimensional geometry. 
They propagate along the flux tube legs in four isolated fronts, whereas
in the two-dimensional Petschek model, they form a continuous,
stationary pair of V-shaped fronts.

We find that the
cross sections of these reconnected flux tubes appear
as teardrop shaped bundles of flux propagating away
from the reconnection site.  Based on this, we argue that 
the descending coronal voids seen by Yohkoh SXT, LASCO, and 
TRACE are reconnected flux tubes descending 
from a flare site in the high corona, for example after a 
coronal mass ejection.  In this model, these flux tubes would then 
settle into equilibrium in the low corona, forming an 
arcade of post-flare coronal loops.  

\end{abstract}

\keywords{Magnetic reconnection---MHD---Sun: flares---Sun: CMEs}

\section{INTRODUCTION}
Coronal mass ejections (CMEs), sudden eruptions of coronal
plasma and magnetic field into the interplanetary medium,
are one of the most dynamic phenomena in the solar corona.
When these ejections collide with the Earth's magnetosphere,
they can have dramatic effects on the Earth's space weather.
Even if they miss the Earth, these CMEs are an important driver
of space weather through the magnetic reconnection and flaring
that is driven in the corona in their wake. Particles
accelerated during the flares associated with CMEs can damage
satellites and create radiation hazards for astronauts, and significant
EUV and X-ray emission is generated by the coronal plasma heated by
these flares.

Theories for the cause and form of CMEs vary 
\citep[see, e.g., reviews by][]{Forbes2000,Klimchuk2001,Low2001, 
Lin2003}, but in all
cases the CME is a magnetically dominated structure, and magnetic
energy release plays a prominent role in its eruption.  In these
models, magnetic field lines, carrying coronal plasma
with them, bow out from the low corona to form the CME, as 
in the simulation shown in Figure 1 \citep{MacNeice2004}.

A critical step
of the process is the reconnection of magnetic field in the wake
of the CME.  Prior to such reconnection
the inward and outward directed field lines in the two
legs of the CME are pinched together to form a current sheet.
When the field reconnects across this sheet, as shown in the right
hand panel of Figure 1, the CME is untethered from the
Sun while the field on the sunward end of the sheet forms post-eruption
coronal loops.  As more of the field reconnects, these post-eruption
coronal loops build up on top of each other to higher and
higher altitudes, and the footpoints of each new set of loops are more and
more widely separated.  This classic two-dimensional flare model 
\citep[see, e.g.,][]{Carmichael1964,Sturrock1966,Hirayama1974,Kopp1976} 
explains both the
appearance of progressively higher coronal loops in a post-CME reconnection
event, and the appearance of progressively more widely separated H$\alpha$
footpoints of the loops.

The basic process of reconnection across a current sheet was first
studied in two-dimensional, steady-state models which assumed uniform
resistivity \citep{Sweet1958,Parker1957}.  The reconnection rates in
such models were usually too low to explain the fairly rapid
development observed in CMEs \citep[see, e.g.,][]{Biskamp1986}.
A faster reconnection regime was proposed first by \citet{Petschek1964}; 
it was distinguished by a very small diffusion region with four slow-mode
shocks linked to it.  It turns out such a structure is inconsistent
with spatially uniform diffusion \citep{Kulsrud2001}, but occurs 
naturally whenever the magnetic diffusion coefficient, or any other 
non-ideal field line
transport mechanism, is locally enhanced.  This explains the notable
successes of a number of modified reconnection models
whereby fast reconnection can occur across a
current sheet provided it is spatially localized, as by a locally
enhanced resistivity \citep{Ugai1977,Scholer1987,Erkaev2000}.
The degree of enhancement is not nearly as important as its
localization \citep{Biskamp2001}.

While the hypothesis of localization solves the puzzle of fast
reconnection, it raises new questions about how such a small-scale
phenomenon might couple to the global magnetic field, say of the CME.  A
partial answer was provided by investigations of two-dimensional,
localized, non-steady reconnection beginning with \citet{Semenov1983}
and subsequently developed by \citet{Biernat1987}, \citet{Heyn1996}
and \citet{Nitta2001}.  These models postulate a localized, non-steady
reconnection electric field
on a pre-existing current sheet and solve for the
external response.  The time-dependent solution,
shown in Figure \ref{fig:biernat}, consists of a pair of 
post-reconnection loops
retracting at the Alfv\'en speed.  Some models also include a
small-amplitude fast mode wave propagating radially outward, in advance
of the slow-mode shocks \citep{Semenov1983,Heyn1996,Nitta2002};
others such as \citet{Biernat1987} treat this as a static disturbance
to the up-stream field.

This two-dimensional model demonstrates several respects in which
localized, non-steady reconnection differs from the more well-known
steady-state scenario.  For example, during reconnection
four slow-mode shocks extend from the reconnection site
as in the steady-state Petschek model, but they do not extend to
infinity, as in the steady-state. Rather they 
close around the tips of the unreconnected flux tubes
(see Fig \ref{fig:biernat}).
In addition, when reconnection ceases the flux-tubes
continue to retract, with the teardrop-shaped slow-mode shock 
fronts remaining intact, and a new current sheet forming behind them.  
Even then, the magnetic energy of the system continues to
decrease as the flux tubes retract.  This
means that the energy released by the reconnection (i.e., converted
from magnetic to kinetic and thermal energy by the shocks)
can far exceed the energy actually dissipated (converted directly to
heat by the reconnection electric field).  Finally, the loops 
continue to sweep up mass into their slow-mode shock envelope as 
they retract, thereby growing indefinitely through
snow-plow--like effect \citep{Semenov1998}.

The present work is aimed at generalizing this scenario
to localized, non-steady
reconnection in three dimensions, which we call {\em patchy
reconnection}.  The objective is to characterize the effect on the
global field of reconnection occurring within a small region of a
pre-existing current sheet.
In particular, we solve for
the dynamics of the post-reconnection flux tubes once they have been
created by some reconnection process.  This is a question of
fundamental importance in the study of magnetic reconnection, as well
as a critical element in understanding CMEs.

Observational approaches have been slow to yield this understanding
since post-reconnection fields
are not observable immediately after the CME. Rather,
only the resulting changes in the coronal configuration are observed.
This is because newly forming reconnected loops only become
visible in emission after most of their energy has been released and
they fill up with hot plasma. This stage only occurs as the loops approach
their equilibrium state in the low corona, and so both the initial
reconnection and the majority of the subsequent dynamics are unobserved.
Thus even truly dynamic studies such as those of \citet{Forbes1996} can
only observe the end stage of the loop evolution.
The apparent dynamics which the two-dimensional models explain so well are simply the
gradual brightening of successively higher coronal loops with
successively wider footpoints. These two-dimensional reconnection models can therefore
only compare their predictions with observations of the final equilibrium
of the magnetic field. While they have been quite successful in this
endeavor, the dynamics of the reconnection and even of the loops themselves
remain largely untested.

A newly observed phenomenon exhibiting genuine three-dimensional characteristics
now offers a promising way to probe the coronal dynamics of reconnection:
dark voids, shown in Figure \ref{fig:sheeley}, are observed by TRACE
and Yohkoh SXT descending through the haze of 15MK post-eruption loops in
the corona \citep[see, e.g.,][]{McKenzie1999,Gallagher2002,Innes2003a,
Asai2004, Sheeley2004}.  These appear to be truly dynamic phenomena, 
as they remain coherent during their evolution, with dominant, trackable 
features. In addition, these descending voids are clearly three-dimensional phenomena, 
as their appearance breaks up the apparent two-dimensional symmetry of the 
post-eruption arcade of hot loops in Figure \ref{fig:sheeley}.
A study by \citet{Sheeley2004} even provides evidence that
the tracks of these voids through the high corona map directly to the
tracks of bright loops which subsequently appear lower down in the corona.
The implication, therefore, is that these descending
voids are actually evacuated three-dimensional magnetic loops which have not yet filled
with hot plasma to become visible in emission \citep{McKenzie1999}.

These descending coronal voids resemble the teardrop-shaped slow-mode
shocks from the two-dimensional model of \citet{Biernat1987},
making non-steady reconnection a promising explanation of 
this phenomenon.  But the fact that the observed voids break the 
two-dimensional symmetry and then form into fully three-dimensional 
loops means that they must be a three-dimensional manifestation of this model.  
We therefore hypothesize that 
these descending voids are flux tubes which were generated by
a short duration, three-dimensional patch of reconnection in the post-CME current
sheet.  Their easily trackable dynamics provide a golden opportunity 
for modeling. Our goal is to learn what we can about
these coronal voids by extending the non-steady reconnection theory into three dimensions. 

In this manuscript, we focus, as in the corresponding two-dimensional 
steady-state model of
\citet{Petschek1964} and the two-dimensional non-steady model of 
\citet{Biernat1987}, on the dynamics of reconnected fields rather than
the mechanism which causes reconnection itself.
Sections 2 and 3 explore a simplified scenario
whereby reconnection and post reconnection dynamics occur in distinct
phases.  In \S 2 the first phase of brief, high-resistivity reconnection
is shown to create a small pocket of potential 
field in the otherwise undisturbed current sheet.  This forms
two flux tubes bent across the current sheet, which then serve as initial 
conditions for the second, post-reconnection, dynamic phase.  
The dynamical relaxation of these flux tubes in the presence of 
unreconnected flux is developed in \S 3.  This development is done 
first in terms of an ideal thin flux tube model and then modified to 
include additional external effects such as ``added mass''.  Having 
developed these theories, we then turn to three-dimensional 
magnetohydrodynamical (MHD) simulations to study the details of 
these dynamics in more detail
in \S 4.  First, in \S 4.1, we initialize a set of simulations with
the analytically reconnected state derived in \S 2. 
We study how the retraction of the reconnected
flux in these simulations compares with the theory of \S 3.
We then proceed, in \S 4.2, to more realistic simulations in which 
we impose reconnection in a one-dimensional current sheet by
increasing the resistivity in a sphere on the
sheet for a short time. In \S 5, we analyze the added mass effect
on the simulated flux tubes, estimating the importance of
this effect. Finally in \S 6 we summarize our results and
briefly discuss their implications for explaining the form and dynamics 
of post-flare coronal voids and arcades.

\section{THE MAGNETIC RECONNECTION EPISODE}

For this study, we assume that the current sheet in which the 
reconnection occurs
is one-dimensional, leaving more complex configurations for
later studies. For the analytical derivation of the reconnected
state, we assume the initial magnetic field on either
side of the current sheet is uniform, with magnitude $B_0$, 
and that the current sheet is formed by a tangential discontinuity 
in the magnetic field at the $z=0$ plane.  The 
discontinuity has half-angle $\zeta$, making the explicit form of the 
field
\be\label{eq:B_init}
  \bvec ~=~ \yhat B_0\cos\zeta + \xhat B_0\sgn(z)\sin\zeta,
\ee
where $\sgn(z) = \pm1$ is discontinuous at $z=0$.
This is an equilibrium, provided the resistivity is exactly zero.

As a first step in our study we will consider a hypothetical scenario
where the reconnection itself and the post-reconnection dynamics occur 
in distinct phases.  That is to say a magnetic reconneciton episode (MRE)
occurs in which the resistivity is greatly enhanced within a sphere
of radius $\delta$ over a very brief period.  To accomplish the desired
separation we take the period to be so brief that the plasma remains 
stationary both inside and outside the sphere.  The resistivity 
is so greatly enhanced that at the completion of the episode the
field inside the sphere is completely current free: ${\bf B}=\nabla\chi$.
The potential $\chi$ is harmonic within the sphere, and satisfies Neumann
conditions to match the radial component of the field of Equation
(\ref{eq:B_init}).  The general solution in a sphere centered at
$\xvec_0=(x_0,y_0,z_0)={\bf 0}$ can be written in terms of a normalized potential
$\chi_0(\xvec)$
\be\label{eq:chi-general}
  \chi(\xvec) ~=~ B_0\cos(\zeta)\,y ~+~ 
  B_0\sin(\zeta)\delta~\chi_0\left(\frac{\xvec}{\delta}\right).
\ee
The normalized potential is harmonic within the unit sphere and matches
the  radial component of $\bvec/B_0=\sgn(z)\xhat$ at the outer boundary
(i.e., it solves the $\zeta=\pi/2$ case).

\subsection{Two-Dimensional Case}

To match previous two-dimensional reconnection studies 
\citep[see, e.g.,][]{Petschek1964, Biernat1987, Nitta2001, 
Hirose2001, Birn2002, Huba2005}
we first consider the case with translational symmetry along $\yhat$.
With this symmetry, the reconnection region is a cylinder with axis at
$x = 0$. Writing
\be
 B_y = B_0\cos\zeta,
\ee
\be
 \xhat B_x + \zhat B_z 
     = B_0\sin(\zeta) \delta~\nabla \chi_0 
     = B_0\sin(\zeta) \delta~\yhat \times \nabla A_0,
\ee
we can define the complex function
\be
     F\left(w=\frac{x+iz}{\delta}\right)=A_0-i\chi_0.
\ee
The Cauchy-Riemann equation gives the magnetic field components
from the derivative of $F$:
\be
     B_z + i B_x = -d_{w}F.  
\ee

To find $F$, we expand $\chi_0$ as an infinite series of cylindrical
harmonics in $r=(x^2+z^2)^{1/2}$ and $\phi=\tan^{-1}(z/x)$,
and set its radial derivative at the surface of the cylinder
($r=\delta$) equal to the radial
component of the magnetic field there:
\be
   \left.\frac{\partial\chi_0}{\partial_r} \right\vert_{r=\delta} = 
           \sum_{m=0}^{\infty} m A_m \left(\frac{r}{\delta}\right)^{m-1} \sin(m\phi) =
           \sgn(z)\xhat\cdot{\hat {\bf r}} =
           \cos(\phi)~\sgn(\sin\phi).
\ee
Using the orthogonality relations for $\cos(m\phi)$, we solve for
the constants $A_m$ to give
\be
   \chi_0=\frac{1}{\pi}\sum_{m=0}^{\infty}\frac{1}{m^2-1/4}
                    \left(\frac{r}{\delta}\right)^{2m}\sin(2m\phi).
\ee
Using $\chi_0 = -Im(F)$, we replace $(r/\delta)^{2m}\sin(2m\phi)$ with
$w^{2m}=(e^{i\phi}r/\delta)^{2m}$ to find:
\be\label{eq:2dpotential}
  F(w) ~=~ {1\over\pi}\, {1 - w^2 \over w}~
  \ln \left( {1 + w \over 1 - w} \right) ~~,
\ee
where we have summed over the infinite series.
This function is analytic within the disk $|w| < 1$.  

Contours of the flux function $A_0(x,z)$ are shown in Figure 
\ref{fig:2d}, giving the projections of field
lines onto the $x-z$ plane.  The field lines
form four different flux systems.  Two of these systems
are layers bent back upon themselves around the remaining current
sheets.  We refer to these two systems as
{\em reconnected flux tubes}, because they consist of field lines which 
have changed their topology so that they now connect from one flux system 
(at $z>0$) to a distinct second flux system (at $z<0$).  On either side of
these bent layers are regions of flux which remain entirely in a single
flux system. Some of these fieldines have been modified by the resistivity
and, if the guide field $B_y$ is nonzero, have also changed their connections.
These could therefore also be referred to as reconnected field lines
\citep[see, e.g.,][]{HornigN2005},  but for this paper, we
only refer to the field lines which cross between the two flux regions
as reconnected field lines.

The reconnected flux per unit length in $y$ in each reconnected
flux tube is given by the flux crossing the $z=0$ plane between 
$x=0$ and $x=\delta$. This is $A_0(w=0)-A_0(w=1) = Re[F(0)]-Re[F(1)] 
= 2/\pi$. 
Thus the reconnected flux per unit length is
\be
  \Delta\Phi ~=~ \frac{2}{\pi}B_0\delta\sin\zeta.
\ee
By the same reasoning, the flux entering the reconnection
region per unit length is $Re[F(i)]-Re[F(1)]=1$, or
$\Phi_{2D}=B_0\delta\sin\zeta$, and so
a fraction $2/\pi$ of the flux entering the
reconnection region reconnects from one side of the current
sheet to the other.  The remaining flux is distorted 
by the diffusion, but is not topologically changed,
in that it does not cross the current sheet.
In this case, the guide field flux does not
contribute to the total flux entering the reconnection cylinder, as it 
lies parallel to this cylinder, and so never crosses its surface.

The current which was formerly within the reconnection region 
$-\delta < x < \delta$, has been diffused to the cylinder's surface
where it forms a surface current (see Fig. \ref{fig:2d}).  This is
not in equilibrium; there is a force at the cylinder surface, which
ultimately drives plasma motion.

The magnetic energy dissipated, per unit length in $y$, by the resistivity 
in the cylinder is
\be \label{eq:en_2d}
 \frac{\Delta E_{d,2D}}{E_{0,2D}}= 
       1-\frac{1}{\pi}\int |\nabla\chi_0|^2 r dr d\phi
     =1-\frac{2}{\pi^2}\sum_{m=1}^{\infty} \frac{m}{(m^2-1/4)}\simeq 0.595 ~,
\ee
where $E_{0,2D}=\pi\delta^2B_0^2\sin^2(\zeta)/8\pi$ is the initial
reconnection (non guide field) component of the magnetic energy per unit 
length in $y$.

\subsection{Three-Dimensional Case}

In three dimensions, the MRE occurs within a sphere of radius
$\delta$. Expanding $\chi_0$
in spherical harmonics, matching $\partial \chi_0 /\partial r$
on the surface of the sphere to the radial component of the external 
magnetic field there, and using the orthogonality relations for the
Legendre Polynomials $P_{n}(\cos\theta)$, gives
\be \label{eq:3d-chi}
  \chi_0\left(\frac{r}{\delta},\theta,\phi\right) 
  ~=~ \cos\phi\sum_{n=1}^{\infty}
  {(4n+1)P_{2n}(0) \over 4n(n+1)(2n-1)}\,
  \left(\frac{r}{\delta}\right)^{2n}\, P^1_{2n}(\cos\theta) ~~,
\ee
with $r=(x^2+y^2+z^2)^{1/2}$, $\tan\phi=y/x$, and $\cos\theta=z/r$.
Here $P^m_{n}(\cos\theta)$ is the associated Legendre function.
Using this in the expression for the full field (Eq. \ref{eq:chi-general})
allows field lines to be traced from a
plane $y = y_a < -\delta$ 
into and out of the the reconnection region, and
on to a plane $y = y_b > \delta$, as shown in Figure \ref{fig:3dfl}.
Field lines which encounter the reconnection sphere compose two
semi-cylinders, on opposite sides of $z=0$.  Beginning at $y=y_a$ on
$z>0$ the field lines break into a reconnected tube which
crosses over to $z<0$, and a distorted but unreconnected tube which 
remains within $z>0$.  The separatrix consists of field lines which map to
$z=0$ at $r=\delta$.

All field lines in the reconnected flux tube must cross $z=0$ within
the sphere.  In the normalized field, the magnetic flux crossing the
half-plane $z=0, x>0$ is 
\be
  -\int\limits_{-\pi/2}^{\pi/2}d\phi \int\limits_0^{\delta} dr\,
  \left.{\partial \chi_0\over \partial \theta}
  \right\vert_{\theta=\pi/2}  ~=~
  \sum_{n=1}^{\infty}{4n+1 \over (n+1)(2n-1)}\delta[ P_{2n}(0) ]^2
  ~\simeq~\delta~~.
\ee
The reconnected flux within the tube which crosses from $z<0$ to $z>0$
is therefore
\be
  \Phi_{tube} = B_0\delta^2\sin\zeta = \frac{\Phi_0}{\pi}\sin\zeta ~,
  \label{eq:phi_tube}
\ee
where $\Phi_0=B_0\pi\delta^2$ is the flux incident on the sphere.
The second reconnected flux tube, shown in Figure \ref{fig:3dfl},
contains the same flux, so $2\sin(\zeta)/\pi$ 
of the flux incident on the sphere has reconnected. 
In the limit $\zeta=\pi/2$, this reduces to the two
dimensional limit. 

The magnetic energy dissipated by the resistivity is
\be\label{eq:en_3d}
  \frac{\Delta E_d}{E_0} = 1 - \frac{\pi}{4}\sum_{n=1}^{\infty}
  \frac{(2n+1)(4n+1)}{(n+1)^2(2n-1)^2}[P_l(0)]^2
         \simeq 0.602~, 
\ee
where $E_0=\sin^2(\zeta) B_0^2 \delta^3/6$ is the 
reconnection component of the initial magnetic energy in the sphere.
This is very close to the fraction of energy released by dissipation
in the two-dimensional case.

\section{POST-RECONNECTION EVOLUTION}

\subsection{Thin Flux Tube Equations}

Prior to the reconnection episode the magnetic field is in mechanical
equilibrium, since the magnetic field on both sides of the current
sheet is uniform.  Following the MRE there are two post-reconnection
flux tubes with sharp bends; these are clearly out of equilibrium.
The object of the present work is to model the dynamical evolution of
these structures, to ascertain how the localized reconnection affects
the global field.

Since the reconnection has created flux tubes, one model for their
evolution can be sought in the dynamical equations for
isolated, thin flux tubes proposed by \citet{Spruit1981}.  
Such models have
traditionally been applied to cases with large plasma $\beta$
where an isolated flux tube is confined by the pressure of
unmagnetized surroundings.  The underlying approach of considering
the net forces acting on segments of a tube should, however, be equally
applicable to the present situation, where the flux tube is
surrounded by magnetic field, possibly at very small $\beta$.

We characterize one flux tube by its axis $\rvec$,
parameterized by its arc-length $\ell$.
The tube encloses a total axial magnetic flux, $\Phi_{tube}$, in the form of
field with strength $B(\ell)$.  The field strength is determined by
pressure balance across the tube.
In low-$\beta$, pressure balance requires that
the magnetic field strength within the flux tube match that in
the layers between which it is sandwiched: $B(\ell)=B_0$.
This is the primary respect in which the current case differs from the
more conventional, high-$\beta$ case.

A material point, $\rvec(\ell)$, accelerates due to the net forces
acting on it.  If the only force is the imbalance of
magnetic tension, resulting from curvature of the field, the governing
equation is
\be\label{eq:fteq}
  \left.{\partial\vvec\over\partial t}\right\vert_{\perp} ~=~
  {B^2\over4\pi\rho}\,{\partial\tvec\over\partial\ell} ~~,
\ee
where $\tvec\equiv\partial\rvec/\partial\ell$ is the tangent vector.
Writing this in terms of $\rvec$ gives a wave equation:
\be\label{eq:ftubewave}
  \left.{\partial^2\rvec\over\partial t^2}\right\vert_{\perp} ~=~
  v_A^2{\partial^2\rvec\over\partial\ell^2} ~~,
\ee
where $v_A^2\equiv B^2/(4\pi\rho)$ is the Alfv\'en speed.

The standard solution to this equation is $\rvec=\rvec(\ell\pm v_At)$.
In general, this is the superposition of two waves traveling
in opposite directions.
As an example, we take the shape shown in Figure \ref{fig:tft}{\it a},
\be
 {\bf r}(l,0)={\bf r_0}
     -\xhat |\ell| \sin{\zeta}
     +\yhat \ell \cos{\zeta},
\ee
as an initial state.
The solution to Equation (\ref{eq:ftubewave})
which has this shape and zero velocity at $t=0$ is
\be\label{eq:xsoln}
 {\bf r}(\ell,t)={\bf r_0}
     -\xhat\frac{| \ell + v_A t|+| \ell -v_A t|}{2} \sin{\zeta}
     +\yhat\frac{ \ell + v_A t+ \ell -v_A t}{2} \cos{\zeta}.
\ee
Taking the derivative of Equation (\ref{eq:xsoln})
with respect to time gives the velocity
\be
 {\bf v}= \xhat{v_A\over 2} \sin{\zeta}
          \left( -\frac{ \ell+v_A t}{| \ell + v_A t|}
          +\frac{ \ell-v_A t}{| \ell - v_A t|}\right).
\ee
Taking the derivative of Equation (\ref{eq:xsoln})
with respect to $\ell$ gives the tangent vector
\be
 {\bf \hat t}=-\xhat{\rho_0\over 2}\left(
          \frac{ \ell+v_A t}{| \ell + v_A t|}
          +\frac{ \ell-v_A t}{| \ell - v_A t|}\right)
           \sin{\zeta}
          +\yhat\cos{\zeta}.
\ee
Together, these two satisfy Equation (\ref{eq:fteq}), generating
a set of straight segments, shown in Figure \ref{fig:tft}{\it a}-{\it c}.
The central segment moves at velocity $\vvec=-\xhat v_{A,0}\sin\zeta$, 
while the remaining two segments
of the tube remain fixed.  The corners between segments move at a
constant speed $|d\ell/dt|=v_{A,0}$, where $v_{A,0}$ is
the Alfv\'en speed in the stationary segments.
In the case where the initial profile forms the hat shape of Figure
\ref{fig:tft}{\it d},
as many field lines do in the analytical potential state
derived above, the dynamic flux tube consist of five straight segments,
as shown in Figure \ref{fig:tft}{\it e}-{\it f}.

This analytic solution is a generalization of the two-dimensional
solution of \citet{Biernat1987}, but differs in several key respects.
The slow-mode shocks correspond, in this case, to the bends
propagating at the Alfv\'en speed.  These are not simple shocks,
owing to the influence of unshocked external field on the tube.  It is for
that reason that the field strength is unchanged by the bend at low-$\beta$. 
It is a peculiarity of the two-dimensional case that makes four different
shocks, propagating along the four legs of the post-reconnection flux
tubes,  merge pairwise into two teardrops.  Lacking this peculiarity,
the three-dimensional case has its full compliment of four separate
shocks.

Another important feature of this three-dimensional solution is the 
absence of snow-plowing.  In the two-dimensional model the retracting flux 
tube sweeps up mass which 
must be accommodated by an ever-growing bubble (see Fig.
\ref{fig:biernat}).  In the
three-dimensional case, however, the bends move apart to create an
expanding region in which to accommodate the new mass.  
The moving segment between the bends has been shortened by a factor 
$\cos\zeta$, and so the material on that segment has been compressed to 
a constant density of $\rho=\rho_0/\cos\zeta$,
while the tube's cross section is unchanged (due
to the assumed constancy of $B$).  A version of the
two-dimensional solution might be recovered by setting $\zeta=\pi/2$.
However, the divergence in the post-shock density is a signal that the
general solution does not include the possibility of an expanding
bubble or cross section.

\subsection{Energy Release}

When the reconnected flux tube retracts, it releases magnetic energy
by shortening the length of its field lines. At time $t$ from the
reconnection episode, the shock has moved a distance $\lambda(t)=v_A t$
along the flux tube in both directions. Or, if the shock has
reached the boundary by time $t$, $\lambda$ is simply the original
length of the flux tube from the boundary to the reconnection site. 
The segment of the flux tube exposed to the shocks by this time has an 
initial length of $2\lambda$, but is shortened by the passage of the shocks 
to $2\lambda\cos\zeta$ (see Fig. \ref{fig:tft}).  
This means that the magnetic energy per reconnected 
tube has been decreased by $2\Phi_{tube} B_0\lambda(1-\cos\zeta)/8\pi$.
This energy is converted into
the kinetic and thermal energy of the moving flux tube segment, in a
manner analogous to the conversion of magnetic energy by
the slow-mode shocks in the solution of \citet{Biernat1987}.
Normalizing energy such that $E^{\prime}\equiv 8\pi E/(B_0\Phi_0)$,
the energy released by the retraction of both reconnected tubes is
\be
 \Delta E_{ft}^{\prime}=\frac{4}{\pi}\lambda\sin(\zeta)(1-\cos\zeta),
\ee
where we have set $\Phi_{tube}=\Phi_0\sin(\zeta)/\pi$ (Eq. \ref{eq:phi_tube}).
For a small reconnection region, $\delta\ll\lambda$, this is
much larger than the energy released in the diffusive
reconnection event itself:
$\Delta E^{\prime}_{rec}\sim 0.8\delta\sin^2(\zeta)$
(Eq. \ref{eq:en_3d}).

In addition, the ambient, unreconnected field also
releases energy due to the reconnection. When the pair of flux tubes
retract from the reconnection region, they vacate a volume
of length $2\lambda(1-\cos\zeta)$ and area $2\delta^2\sin\zeta$.
Unreconnected field then expands slightly to fill the void, meaning
that it must weaken slightly. This external field  
has a pre-expansion energy of 
\be
  E_{ext}=\frac{B_0\Phi_{ext}2\lambda(1-\cos\zeta)}{8\pi}.
\ee
Here $\Phi_{ext}$ includes all the flux not in the two reconnected
tubes, possibly infinite in extent:
\be\label{eq:flux_init}
 \Phi_{ext}=B_0(A_{full}-2\delta^2\sin\zeta),
\ee
where $A_{full}$ is the full area of interest, e.g., the simulation
area or a significant fraction of the corona.
When the reconnected tube retracts, the external
field expands to fill the vacated volume, decreasing
in strength to $B_0-b$ and expanding in area to $A_{full}$
while keeping the same flux.
Setting the initial flux (Eq. \ref{eq:flux_init}) equal
to the final flux ( $[B_0-b]A_{full}$), we can
solve for the change in field strength:
\be
 b=\frac{B_0A_{full}-B_0(A_{full}-2\delta^2\sin\zeta)}{A_{full}}
  =\frac{2B_0\delta^2\sin\zeta}{A_{full}}.
\ee
The energy released by this field is then
\be
 \Delta E_{ext}=
  \frac{2b\Phi_{ext}\lambda(1-\cos\zeta)}{8\pi}=
  \frac{4B_0\delta^2\sin\zeta}{A_{full}}
   \frac{B_0(A_{full}-2\delta^2\sin\zeta)\lambda(1-\cos\zeta)}{8\pi}. 
\ee
Taking the limit $A_{full}\gg \delta^2$,
we find that the energy released in the external field is
\be
 \Delta E_{ext}^{\prime}=\frac{4}{\pi}\lambda\sin(\zeta)(1-\cos\zeta),
\ee
which, remarkably, is the same as the energy released 
in the flux tube itself as it retracts.

\subsection{Inclusion of External Forces}

The foregoing analysis provides an idealized
flux tube evolution following reconnection.
To obtain the clean analytic form it assumed that the only force
acting on a tube segment is the magnetic tension of the tube
itself.  There are in general other forces due to the surrounding
plasma and un-reconnected field.  These forces modify the flux
tube dynamics in complicated ways.

To estimate these additional effects, we assume that the solution 
retains the general form of the idealized one: 
bends propagate at the Alfv\'en speed, creating a post-reconnection segment
moving with a mean velocity (i.e., center-of-mass velocity) of $V$.
The net forces on the post-reconnection segment, or simply ``segment'', 
increase its momentum.  If the segment achieves a steady velocity then
\be
  F_{\rm net} = {dP\over dt} \simeq V\frac{dM}{dt},
\ee
where $M$ is the total mass of the segment.  As the bends move
at speed $v_{\rm A}$ along the two tube legs they sweep up mass at a
rate
\be
  \frac{dM}{dt} =2\rho_0 v_{\rm A} \frac{\Phi_{tube}}{B_0},
\ee
where $\rho_0$ is the density of the ambient plasma, and $\Phi_{tube}/B_0$ is
the cross sectional area of the tube.

The magnetic tension force along a flux tube, ${\bf F}=\Phi_{tube}{\bf B}/4\pi$,
is directed along the axis.   The contribution in the direction of motion
from a single leg is
\be
   F_{\rm leg} = \frac{1}{4\pi}\Phi_{tube} B_{\perp},
\ee
where $B_{\perp}=B_0\sin\zeta$ is the component of the magnetic
field along the direction of motion.

There is also a drag force on the segment, which matches the
force the segment must exert on the external plasma.  It exerts
this force in order to deform the un-reconnected flux and to
accelerate a wake of entrained fluid.  We assume that the work
done by these forces is a multiple $X\ge0$ of the kinetic energy of the
segment itself. In terms of this factor the net force can be written
\be
  F_{\rm net} = \frac{2}{4\pi}\Phi_{tube} B_{\perp}
  - X {dP\over dt},
\ee
and Newton's law becomes
\be
   \frac{2}{4\pi} \Phi_{tube} B_{\perp} =  2 (X+1) V \rho_0 v_{\rm A} 
                                          \frac{\Phi_{tube}}{B}.
\ee
Solving for the velocity of the segment's center of mass gives
\be\label{eq:v_tube}
  V = \frac{B B_{\perp}}{4\pi\rho v_{\rm A} (X+1)} =
  \frac{v_{\rm A,\perp}}{X+1}.
\ee
Thus even as the bends propagate along this direction at
$v_{\rm A,\perp}$, the center of mass of the segment follows behind them 
at a fraction, $1/(X+1)$, of that speed due to the drag it experiences from
its surroundings.  This would naturally cause the segment to become
arched, but otherwise the behavior would resemble that of the idealized
solution.

If the principle external force is one of drag, we can estimate
the factor $X$ using the ratio of kinetic energies of the reconnected
flux to the un-reconnected flux.  Assuming these two components
move at similar speed,
\be\label{eq:x}
X = \frac{K_{unrec}}{K_{rec}}.
\ee
Inserting this expression for $X$ into Equation (\ref{eq:v_tube}) then
gives us the expected velocity of the tubes.
In \S 5, we calculate this from our simulations to
find out how important this added mass effect is.

\section{SIMULATIONS}

We now use
three-dimensional MHD simulations to test the predictions derived
above for the dynamics of isolated magnetic reconnection events. 
We take these simulations in two stages. The first set of 
simulations reproduce the idealized two-stage process of reconnection
and relaxation. These use the reconnected state 
derived in \S 2.2 as the initial condition. The initial
current sheet therefore has a sphere of potential field
on it, as defined by Equation (\ref{eq:3d-chi}), and only a small
amount of reconnection take place dynamically,
due to the background resistivity of $\delta v_A/\eta_0=200$.
The second set of simulations consider a more realistic case
where reconnection and relaxation occur together. These start
with an unreconnected current sheet and impose a sphere of high resistivity,
$\delta v_A/\eta=2$, on the sheet for a short, but finite, time. Once the 
reconnected flux equals the reconnected flux in the initial state
of the potential simulations, this high resistivity region is 
turned off.

As these simulations have a finite grid size, we cannot
simulate the infinitely thin current sheet of Equation
(\ref{eq:B_init}). Rather, we simulate a finite width Harris current sheet,
with a guide field component added.
The magnetic field of the current sheet, shown in Figure 
\ref{fig:bbla_b}{\it a}, is then given by 
\be \label{eq:bwithdip}
 {\bf B}=B_0 \sin(\zeta) \tanh\left(\frac{z}{l}\right)\xhat+
         B_0 \cos(\zeta) \tanh\left(\frac{|z|}{l}\right)\yhat.
\ee
In this equilibrium, the guide field $B_y$ drops to zero at the center
of the sheet.  We choose this form to represent the scenario where
the two legs of a CME collide to form the current sheet, with
no flux initially in between them. This configuration naturally
generates a current sheet with a thin, zero field region between the
two collided flux systems. While we focus primarily
on this form of the current sheet, for comparison we 
also simulate two other guide field profiles. The first 
is where $B_y$ is constant across the current sheet: 
$B_y=B_0 \cos{\zeta}$.  The second is a low-$\beta$ equilibrium,
where $|{\bf B}|$ is uniform across the current sheet, requiring
that $B_y=(B_0^2-B_x^2)^{1/2}$. 

In all simulations, we set the current
sheet half-width to be the same as the pixel size: $l/L=1/64$,
where the simulation cube is $2L$ on a side, and the
simulations are run at $128^3$ resolution. The radius of
the reconnection sphere is $\delta/L=\pi/16$. The pressure
is set such that $p+B^2/(8\pi)=constant$. Far from the current
sheet, the pressure is $p=p_0=20/3$ in units where 
$B_0/(8\pi)^{1/2}=1$ for the high-$\beta$ simulations,
and $B_0/(8\pi)^{1/2}=4$ for the low-$\beta$ simulations.
This gives $\beta=8\pi p_0/B_0^2=20/3$ and $5/12$, respectively. 
The density is set to be initially uniform everywhere.

We use the MHD code CRUNCH3D \citep[see, e.g.,][]{Dahlburg1995}
to carry out these simulations.
This code solves the compressible, visco-resistive MHD equations, in the 
form presented in \citet{LintonA2005}, with uniform viscosity,
but non-uniform resistivity. The code is pseudo-spectral, and therefore
has periodic boundary conditions.

\subsection{Potential Reconnected Field}

\subsubsection{Simulations}

For the first set of simulations, the initial state has 
a sphere of potential, reconnected field imposed on the current 
sheet, as shown in Figures \ref{fig:bbla_b}{\it a} and \ref{fig:bbla_fl}{\it a}.
Figure \ref{fig:bbla_b} is a $y=0$ cut through the simulation,
showing the magnetic field in the plane as the vectors, and the
field perpendicular to the plane (the guide field) as the
greyscale. Figure \ref{fig:bbla_b}{\it a} shows how the addition of 
the potential sphere to the current sheet creates a set of 
reconnected fields, which form an x-point 
in the midplane of the sphere. These reconnected fields are well
out of equilibrium, as there is no force balancing their tension.
In Figures \ref{fig:bbla_b}{\it b} and {\it c}, 
these fields therefore quickly pull away from the reconnection region, 
forming a pair of teardrop shapes. These shapes are fully
formed by Figure \ref{fig:bbla_b}{\it c}, meaning that all of
the initially reconnected flux has retracted from the reconnection
sphere into the teardrop shapes.
From then, the reconnected fields
simply pull away from the center of the simulation, approximately
keeping the same teardrop shape throughout. This dynamic
resembles the two-dimensional analytic dynamics of \citet{Biernat1987}
(Fig. \ref{fig:biernat}). 

There is, however, a fundamental difference between the three-dimensional
reconnected field shown in Figure \ref{fig:bbla_b} and the two-dimensional 
reconnected field shown in Figure \ref{fig:biernat}. While the envelopes of
the teardrop shapes are slow-mode shocks in two dimensions, this is not
the case in three dimensions. To see these shocks, one must look
at the field lines in a three-dimensional figure, such as Figure \ref{fig:bbla_fl}.
Here, a set of Lagrangian trace particles are followed dynamically during the 
simulation, and field lines are traced from these particles and plotted
at the same times as the slices of Figure \ref{fig:bbla_b}.
The trace particles used for this figure initially lie within
the reconnected sphere along the $y=z=0$ axis.  
These field lines are shown from the $-\zhat$ axis,
perpendicular to the view shown in Figure \ref{fig:bbla_b}. 
The $y=0$ plane of Figure \ref{fig:bbla_b} is represented in Figure 
\ref{fig:bbla_fl}{\it a} by the
white line bisecting the reconnected field lines.
The color table in this figure is proportional to the field aligned
electric current, as shown by the color bar.

Figure \ref{fig:bbla_fl}{\it a} shows that, in the initial state,
all the field lines from these trace
particles are reconnected. Most of the field lines
take the shape of the upper part of a trapezoid, with one bend in the 
field lines as they enter the potential sphere, and a second bend
as they exit.  The solution for their evolution, given by 
the thin flux tube equations presented in \S 3.1, should 
be the superposition of two oppositely propagating
trapezoids, each of half the magnitude of the initial
trapezoid, as in Figures \ref{fig:tft}{\it d}-{\it f}. 
The two bends thus propagate in both
directions along each field line, giving two pairs
of shocks on each flux tube.  These are the slow-mode 
shocks corresponding to that of \citet{Petschek1964} and 
\citet{Biernat1987}.  One of these pairs
is indicated in Figure \ref{fig:bbla_fl}{\it b}-{\it d} by the white
lines labeled as `slow shocks.' The field lines do not change
their shape significantly before the shock hits them,
but then they bend to a new direction as the shock
passes them.  These shocks quickly propagate out of the two
dimensional plane of Figure
\ref{fig:bbla_b} and therefore do not form the boundary of the
teardrop shapes in this two-dimensional cut,
in contrast to the two-dimensional shocks of Figure \ref{fig:biernat}.

We performed this simulation, with a potential sphere
of reconnection, for angles $\zeta=[1,2,3,4]\pi/8$,
as summarized in Figures \ref{fig:bblsumm}{\it a}-{\it d}, respectively.
Here each panel shows two snapshots from the evolution of a 
simulation.  The top half of each panel shows the initial 
state of the simulation, and the bottom half of each panel shows 
the magnetic field in the middle of the evolution.  
The three simulations with a guide field ({\it a}-{\it c})
show the slow-mode shock structure discussed above. Panel {\it d},
where the guide field is zero, is the closest to the two-dimensional limit,
and therefore shows that the field lines form a bulging teardrop shape 
quite similar to the two-dimensional configuration predicted by \citet{Biernat1987} 
(Fig. \ref{fig:biernat}).  

\subsubsection{Analysis}

According to the thin flux tube theory of \S 3.1,
the reconnected flux tubes should retract from the
reconnection region at the reconnection Alfv\'en speed 
$v_{A\perp}=v_A \sin\zeta$.  
To test this, we measure the speed of the center of the flux tube,
i.e., the part lying along the $y=z=0$ axis. Figure \ref{fig:ht}{\it a}
shows a slice of $B_z$ along the $y=0$ plane, along with a 
white dashed line along the $y=z=0$ axis. We take 
the profile of $B_z$ along this line at successive timesteps 
and stack the them vertically to create the distance versus time plot
of Figure \ref{fig:ht}{\it b}. In this plot, features moving in the
$\xhat$ direction appear as diagonal lines whose slopes give their
$\xhat$ velocity.  To measure the velocity of the center of mass of 
the flux tube,
we fit a gaussian to the x-profile of positive $B_z$ along this line 
on the $x<0$ side of the simulation at each timestep.
As an example, the centroid of this gaussian is plotted for 
$tv_{A\perp}/L=0.658$ in
Figure \ref{fig:ht}{\it a} as the vertical white line.  This centroid 
is then plotted as a function of time in Figure \ref{fig:ht}{\it b}
as the solid white line, while the $\pm 2\sigma$ level of the gaussian
are plotted as the two dotted white lines. We then find the 
linear least squares fit to this center-of-mass line from the time 
when the teardrops are first fully formed  ($tv_{A\perp}/L=0.27$) until the 
flux tube is about to hit the edge of the simulation ($tv_{A\perp}/L=1.4$). 
The fit, plotted for the appropriate time range as the dashed white line
in Figure \ref{fig:ht}{\it b},
gives a velocity of $v_x=.406 v_{A_\perp}$.

The asterisks in Figure \ref{fig:velocity}{\it a}
show the measured velocities of the potential reconnection flux 
tubes for the four angles simulated. The tube speed is 
consistently lower than the reconnection Alfv\'en speed, $v_{A\perp}$, 
but it approaches this speed as the angle between the reconnecting 
fields increases.  In contrast, the slow-mode shock propagates at the 
full Alfv\'en speed, as expected. For $\zeta=\pi/4$,
a distance versus time plot with distance along direction
of shock propagation, i.e., along the $x=y$ diagonal,
gives a shock propagation speed of $.951v_A$.
As the thin flux tube analysis of \S 2
predicts that the tube speed should be $v_{A\perp}$
if the shock speed is $v_A$, it is clear that there is a drag force, such
as the added mass effect, on the tubes.  
In addition to the tube speed being slower than 
predicted, a second indication of this drag force
is given by the fact that the flux tubes of Figure \ref{fig:bbla_fl}
do not form straight segments between the shocks, but rather
form curved segments.  

Figure \ref{fig:ht} shows an additional feature of the three-dimensional reconnection
which may provide an answer to the source of the drag force:
the reconnected tubes are sweeping up unreconnected field which 
lies across their path. This field lies on or close to the
current sheet on $z=constant$ planes, in contrast
to the reconnected field which crosses the $z=0$ current sheet plane 
from one set of flux to the other set. The two sets of field are therefore 
topologically entwined, and wrap around each other rather than
flowing over each other. The fact that the reconnected field
crosses the $z=0$ plane in one direction forces the unreconnected
field it sweeps up to warp such that it crosses $z=0$ in the
other direction. This warped field can be seen in
Figure \ref{fig:ht}{\it a}.  The reconnected field forms the
classic teardrop shape, with negative $B_z$ on the $x<0$
side and positive $B_z$ on the $x>0$ side. But in front
of these teardrop shapes the unreconnected, warped
fields are visible as a layer of $B_z$ of the opposite sign.

The second set of simulations we performed studied the effect of having
a stronger guide field in the current sheet. In this case, a magnetic
equilibrium was simulated with $\beta = 20/3$ as before,
but with a uniform guide field, rather than a guide field which
drops toward zero in the current sheet.  If the tubes are slowed
down in part because they drag the guide field in the current
sheet along with them, then the flow in this second
set of simulations should be slower. Indeed we
found that the velocities are slower: $34\%$ for $\zeta=\pi/8$,
$16\%$ for $\zeta=2\pi/8$, and $6\%$ for $\zeta=3\pi/8$.
Both simulations were exactly the same for $4\pi/8$ because
at that angle the guide field is zero everywhere. Thus,
we find that the drag effect becomes stronger as the guide 
field in the current sheet gets stronger.

\subsection{Resistively Reconnected Field}

The potential reconnection simulations of \S 4.1 allowed us to study 
the dynamics of reconnected fields without worrying about
the dynamics of the reconnection itself. Having analyzed this
simpler situation, we now proceed to study both the reconnection
and the subsequent dynamics in the same simulation.
Here we start with the same unreconnected state given by
Equation (\ref{eq:bwithdip}). But 
instead of inserting the potential field of Equation (\ref{eq:3d-chi})
into the reconnection sphere, we leave the initial field unreconnected,
and increase the resistivity in the sphere ($r\le\delta$) to
\be
 \eta=\eta_0(1+99 e^{-r^2/\delta^2})
\ee
where $\eta_0$ is the background resistivity, $\eta_0/\delta v_A=1/200$.
The resistivity stays at this high level until the amount of dynamically 
reconnected flux equals that which is initially reconnected in the
equivalent potential-sphere configuration, and then
it is decreased back to $\eta_0$.

\subsubsection{Two-Dimensional View}

Two-dimensional snapshots of the magnetic field from
such a simulation at $\zeta=2\pi/8$ are shown in 
Figure \ref{fig:neta2f_b}, in the same format as Figure 
\ref{fig:bbla_b}.  Panel {\it a} shows the unperturbed current 
sheet, with contours lines of $\eta$ superimposed to show where 
the resistivity is concentrated. As this sheet will be 
dissipated by the resistive sphere, it is interesting
to follow its structure on the line $x=y=0$, which 
goes through the center of the resistive sphere.
The maximum current strength and the width of the
current at half of its maximum value along this line are 
plotted as a function of time in Figure \ref{fig:sheet}.
This shows that the maximum initial value
of the current in this sheet is $0.7~J_0$, where
$J_0=\nabla\times B=B_0\sin\zeta$ is the prescribed initial
peak current strength. The initial width of the sheet
is $2.7~l$ (where $l$ is one pixel wide).  Note that the 
width of the prescribed profile (Eq. \ref{eq:bwithdip}) is 
$1.8~l$ and so the current sheet is wider
and weaker than prescribed by the initial conditions. 
This is because the code smooths out strongly
peaked features with a raised cosine filter as part of the initialization
routine. 

The effect of the resistive sphere can be seen in Figures
\ref{fig:neta2f_b}{\it b} and {\it c}, where it rapidly diffuses the current
sheet, causing it to thicken. Figure \ref{fig:sheet} shows
that the sheet width jumps to between $12~l$ and $8~l$ while 
the amplitude drops to $\sim 0.12~J_0$.
The diffusion causes the magnetic field to reconnect,
forming a configuration much like
that of the potential configuration. 
This field rapidly pulls away from the reconnection region,
as before. This can be seen in Figure \ref{fig:neta2f_b}{\it c} where
the reconnected flux has started to form into two teardrop
shapes on either side of the reconnection region. At this
time, the diffusive sphere is still active, and so new
flux is being continually added to the reconnected flux
tubes, and the teardrop shapes extend back to the x-point.

The flux reconnected by this diffusion can be estimated by
calculating the integral of $B_z$ crossing the $z=0$ half plane
$x>0$ in the positive direction. 
This is shown in Figure \ref{fig:globals}{\it a}
as the dash-dotted line. At $tv_{A\perp}/L=.34$ as much flux has 
reconnected  ($0.8~\Phi_{tube}$) as is reconnected in the initial 
condition of the equivalent potential reconnection simulation. 
This is smaller than the value, $\Phi_{tube}$, given by the analytical
analysis because the simulated current sheet has finite width $l$, 
and so the initial flux $\Phi_0$ incident on the reconnection sphere is
only $80\%$ of that expected for an infinitely thin current sheet.
At this time, we turn the resistive sphere off. 
Reconnection slows significantly at this point, but does not stop,
due to the low level of background resistivity.

After the resistive sphere is turned off, the teardrops of reconnected
flux continue to move away from the reconnection site,
and the current sheet reforms behind them. Interestingly,
Figure \ref{fig:sheet} shows that the width and amplitude of
the current sheet return to approximately their initial values.

For comparison with this
finite duration magnetic reconnection episode, we simulated
the same interaction but never turned off the resistive
sphere. The result, at the same time as in Figure \ref{fig:neta2f_b}{\it f}
is shown in Figure \ref{fig:neta2f_b}{\it g}. In this case, the x-point 
still exists at $x=0$, and the teardrop shape extends
all the way back to this x-point. 
This shows that the aspect ratio of the flux tube
cross section is strongly influenced by 
the duration of the reconnection episode which creates it.
The leading edge of this flux tube in panel {\it g} has traveled 
slightly further in the same amount of time than the flux
tube with a limited amount of reconnection in panel {\it f}.
This indicates that the accumulation of extra reconnected flux
in the flux tube increases its speed slightly, possibly
by helping to overcome the drag force.
The current sheet characteristics for this steady reconnection
simulation are plotted as the dotted lines in
Figure \ref{fig:sheet}.  This shows
that the width of the sheet 
settles at about $8~l$ and the amplitude stays at 
about $0.12~J_0$ as the reconnection approaches a steady-state.  
The reconnection rate stays the constant through the rest of 
the simulation rather than slowing down as it does when the 
resistivity is turned off. 

The flux tube velocities for these high-$\beta$ resistive
simulations at the four angles $\zeta=[1,2,3,4]\pi/8$ are 
plotted as the triangles in Figure \ref{fig:velocity}{\it a}.
These flux tubes are slower than the flux tubes 
retracting from the potential reconnection region at low angles,
yet they are slightly faster at the higher angles.

Again, we repeated this set of simulations for a flat guide
field and found the same trend as we did for the potential case:
the tubes are slowed by the additional guide field, particularly
as $\zeta$ decreases and the guide field gets stronger.
Here we found that the velocities were $41\%$ slower for $\zeta=\pi/8$,
$17\%$ slower for $\zeta=2\pi/8$, and $12\%$ slower for $\zeta=3\pi/8$.

\subsubsection{Three-Dimensional View}

Figure \ref{fig:neta2f_fl}{\it a} shows the
initial three-dimensional configuration of this simulation, where two sets of straight, 
unreconnected fieldines cross each other at right angles ($\zeta=\pi/4$). 
As for Figure \ref{fig:bbla_fl}, these field lines are taken from Lagrangian 
trace particles, chosen such that they intersect the sphere of high
resistivity at the start of the simulation. 
In this case we show a set which is initially just on either
side of the current sheet.  As the particles
lie in the high resistivity region initially, they
are not frozen onto field lines. They do, however, show
snapshots of the state of the reconnecting field,
and more exactly follow the field line evolution
after they leave the resistive sphere, or after
the resistive sphere is turned off.

Initially, the fieldines are perfectly straight and 
unreconnected. As soon as the simulation starts,
however, the resistive sphere generates reconnection.
The reconnecting field lines `fan' across the current sheet, 
appearing blue because the current sheet has a very high 
parallel electric current. This fanning agrees with the analysis
and simulations of \citet{Pontin2004} and \citet{Pontin2005}.
According to this theory, in three dimensions reconnecting field lines 
continuously change their connectivity while they are actively
reconnecting, rather than undergoing a single connectivity-changing event
as in two-dimensional reconnection.

The resulting reconnected flux tubes
progress in a similar fashion to that of the potentially reconnected
field lines.  It is difficult
to see the slow-mode shocks in this field line figure
because the shock fronts are much more diffuse due to the finite
time it takes for the reconnection to occur as compared
with the instantaneous reconnection in the potential case.  However, 
the shocks can still be followed in diagonal slices of $B_z$ along
the $x=y$ plane. Again, we find that these shocks
travel at close to the full Alfv\'en speed ($\sim 0.8 v_A$ for $\zeta=2\pi/8$). 

The four simulations of this type are summarized in Figure 
\ref{fig:bblsumm}{\it e}-{\it h}.  The top half of
each panel shows the reconnected fields early in their evolution,
while the bottom half shows the tubes in mid-evolution.
The tube dynamics are qualitatively 
similar to those of the potential sphere simulations shown
in Figure \ref{fig:bblsumm}{\it a}-{\it d}. 
A weak shock front can be seen, particularly in the top
half of panels {\it e}-{\it g}, propagating along the flux tubes
as in the potential reconnection simulations.  In addition, the 
arched shapes of the reconnected field lines are qualitatively 
the same in the two types of simulations.

\subsection {Low-$\beta$ Simulations}

The third set of resistive reconnection simulations we
performed are the low-$\beta$ simulations ($\beta=5/12$) 
where the guide field $increases$ in the current sheet to provide 
pressure balance there. We were unable to simulate the potential
simulation at all angles at the same level of viscosity and resistivity
as for the high-$\beta$ simulations, and so we do not discuss
this experiment here. The problem is 
that the fields are very much out of equilibrium
at the start of the potential simulation, and there is less gas pressure
to balance any sudden disturbances at low-$\beta$ than
at high-$\beta$. Therefore the density tends to become
negative in cavitation regions of strong motion, causing
the code to crash. We could solve this problem through the
artificial method of forcing the density to remain larger
than some small value. However, as the resistive reconnection simulations
do not have the same initial non-equilibrium  problem, we focus on 
those simulations instead.

The two-dimensional view of the low-$\beta$, resistive reconnection
simulation at $\zeta=2\pi/8$ is shown in 
Figure \ref{fig:lbeta2d_b}.  This shows that decreasing
$\beta$ from $20/3$ to $5/12$ does not dramatically affect the 
reconnection dynamics.
The current sheet thickens as before when the resistive sphere
is turned on, and two teardrop shaped flux tube cross sections
are formed. The current sheet reforms again once the resistive 
sphere is turned off after Figure \ref{fig:lbeta2d_b}{\it c}. 
The velocities of these flux tubes are lower than in the
high-$\beta$ simulations, as shown by the squares in Figure 
\ref{fig:velocity}{\it a}. 
In addition, the tube velocities are not significantly faster
at high $\zeta$ angles, in contrast to the high-$\beta$ simulations.
Finally, for reference, Figure \ref{fig:lbeta2d_b}{\it g} again shows
a second simulation wherein the resistive sphere is never
turned off, and so the flux tube continues to grow in time.
This tube again travels slightly farther during the same time
than the tube of Figure \ref{fig:lbeta2d_b}{\it f}, which contains
less reconnected flux.

\section{ADDED MASS EFFECT}

The velocities we derive for all of these simulations
pose the question: why are the
tubes moving at a slower speed than predicted by
the thin flux tube model presented above?
A likely possibility is the added mass effect
discussed in \S 3.3.
When a body accelerates through an external
fluid, as the reconnected flux tubes do here,
the external fluid is necessarily also accelerated.
This takes extra energy, and therefore slows
down the flux tube. If this is correct, we should
see that the unreconnected field in the simulation
gains kinetic energy in proportion to the kinetic
energy of the reconnected field. 

To study this, we must
separate unreconnected fields from reconnected fields.
For each time step at which we want to calculate these
kinetic energies, we first trace a volume filling set of field lines,
starting from a uniformly distributed grid of $64^2$
points at the $y=0$ plane. 
To ensure that the field lines
are volume filling, we trace them in both directions
in this plane until they hit the $y=\pm L$ boundaries.
As $B_y$ is greater than zero at all points in the simulation 
(for $\zeta\ne \pi/2$), this routine counts each field line
only once, as no field line doubles back in the $\yhat$ direction.
To fill the whole simulation volume with field lines, this requires 
that we trace field lines out of the simulation
box in the $x$ direction, making use of the periodic
boundary conditions. 
The $\zeta = \pi/2$ case is more involved to study, as
it has no organizing guide field. In this case, we trace 
all field lines from $z>0$ starting at $x=-L$ (the
left hand boundary of the simulation). We trace these field lines
until they reach the $x=0$ plane or the $z=0$ plane.
Assuming symmetry of $B_y$ and $B_x$ and anti-symmetry of
$B_z$ about $x=0$, and symmetry of $B_y$ and $B_z$ and anti-symmetry
of $B_x$ about $z=0$ (the simulation shows both are valid) we can 
recover the whole volume of the simulation without double counting any 
field lines.

Each field line $i$ from the trace is assigned a flux
$d\Phi_i$, given by the field strength $B_i$ at its 
starting point $(x_i,0,z_i)$ multiplied by an area
$4L^2/N$, where $N$ is the total number of field lines traced.
By integrating local quantities along
each field line, we can then find the field line integrated
global quantities, such as volume or kinetic energy.
The field strength varies along a field line, so we
account for the changing area along the field line
by assuming a constant flux: 
$dA_i(\xvec)=d\Phi_i/|{\bf B}(\xvec)|$.
The volume of each field line is then
\be
V_i= \int dA_i dl_i=d\Phi_i\int \frac{dl_i}{|{\bf B}|}.
\ee

As a check of our method, we sum over all N field lines
to find the total volume. We find that this agrees
with the true volume to within $.008\%$ for
$N=64^2$ field lines.
We can then calculate the kinetic energy of each field line:
\be
K_i =d\Phi_i\int \frac{dl_i}{|{\bf B}|} \rho |{\bf v}|^2.
\ee
Summing over the all field lines, we find the kinetic
energy agrees with the true sum to within $3.5\%$ on average.
We then must divide the field lines into reconnected
and unreconnected field lines. Here we take advantage
of the simple geometry of our simulation: 
in the unreconnected state, none of the field lines 
cross the $z=0$ boundary between the differently directed
sets of field lines. Once a field line reconnects, however,
it connects fields on both sides of this plane, and so it
must intersect this plane. Thus, to categorize field lines
as reconnected, we require that they cross the $z=0$ plane
in the negative direction for $x<0$ and in the positive
direction for $x>0$. This rejects field lines which have
wrapped around the reconnected tube but have not reconnected,
as they cross the $z=0$ plane in the opposite direction
(as discussed in \S 4.1.2).

The resulting calculations of kinetic energy for the reconnected 
versus the unreconnected fields are shown in Figure \ref{fig:globals}  
as a function of time for the high-$\beta$ resistive reconnection
simulation at $\zeta=2\pi/8$ (see Figs. \ref{fig:neta2f_b} and 
\ref{fig:neta2f_fl}). 
Here one can see that the kinetic energy of 
the reconnected tubes increases in time, and that the kinetic energy
of the unreconnected field increases in proportion.
Thus, the kinetic energy, and therefore the velocity
of the reconnected tubes must be less than it would
in the absence of an external medium. 

As a check
of our field line sorting algorithm, the dotted
line in Figure \ref{fig:globals}{\it a} shows the
reconnected flux per tube for this simulation, calculated
as half the sum of the flux of all reconnected field lines
in the field line tracing calculation. 
This compares reasonably well with the dash-dotted line,
which shows the reconnected flux measured by summing
all the flux which crosses the $z=0$, $x>0$ plane in the positive
direction.

We find that the unreconnected kinetic energy approximately 
follows the reconnected kinetic energy as in Figure 
\ref{fig:globals}{\it b} for all simulations studied.
To estimate the added mass effect, we calculate
the ratio of the integrated kinetic energy for the unreconnected 
and reconnected field lines:
\be
 X=\frac{\int K_{unrec}dt}{\int K_{rec} dt}.
\ee
where the integrals are taken from the time
the flux tubes are fully formed until they are about
to reach the boundary.
We find that the ratio $X$ increases as $\zeta$ increases. 
This is further evidence that the guide field is acting as a drag on the
flux tubes, as the strength of the guide field decreases
as the angle increases.  

Using the ratio X in Equations \ref{eq:v_tube} and \ref{eq:x}, 
we calculate the expected velocities of the reconnected
tubes due to the added mass effect.  The results are shown
in Figure \ref{fig:velocity}{\it b}.  Comparing these velocities 
with the measured velocities of Figure \ref{fig:velocity}{\it a}, 
we find that, on average, the potential
reconnection tubes move at $82\%$ of this predicted speed, 
the high-$\beta$ resistive reconnection tubes move at $72\%$ of this 
predicted speed, but the low-$\beta$ resistive reconnection tubes 
move at only $49\%$ of this speed. 
Thus we conclude that the added mass
effect accounts for most of the flux tube drag
at high-$\beta$, but there is an additional, significant source of 
drag at low-$\beta$.

\section{CONCLUSIONS}

We have studied non-steady three-dimensional magnetic reconnection
by imposing reconnection in a localized sphere
on a one-dimensional current sheet. We have shown that external 
magnetic field intersecting this sphere of reconnection forms a 
pair of bent, reconnected flux tubes. We have solved analytically for
the post-reconnection evolution of these flux tubes, 
showing that slow-mode shocks propagate as isolated fronts
along the legs of the tubes, releasing magnetic energy and 
accelerating the flux tubes away from the reconnection region. 
These isolated shock fronts are the three-dimensional 
generalization of the two V-shaped shocks of steady two 
dimensional Petschek reconnection \citep{Petschek1964}, and of 
the two teardrop shaped shocks of non-steady two-dimensional 
reconnection \citep{Biernat1987}.

We then studied the evolution of this reconnected magnetic 
field with three-dimensional magnetohydrodynamical 
simulations.  We found that, as predicted, the reconnected field 
quickly retracts from the reconnection sphere, 
accelerating as the slow-mode shock fronts pass by.
This accelerated field forms a pair of three-dimensional, 
arched flux tubes whose cross sections have a distinct 
teardrop shape.  We found that the velocities of these flux tubes
are smaller than the reconnection Alfv\'en speed
predicted by the theory, indicating that some drag
force is slowing them down. We provide evidence that an 
added mass effect, wherein the flux tubes sweep up external
magnetic field and plasma, is largely responsible for this 
drag force.

These three-dimensional reconnected flux tubes are a promising
candidate for explaining observations of descending coronal
voids seen below post-CME flares. Their teardrop shaped 
cross section is similar to the shape of these voids \citep{McKenzie1999}, 
and their three-dimensional arched shape is similar to the shape
of the coronal arcade loops formed by these voids later in their
evolution \citep{Sheeley2004}. We therefore argue that these
coronal voids are flux tubes formed by localized patches of reconnection
higher in the corona.  If this is the case, then by studying the dynamics
of these voids, we can directly study the dynamics of reconnected 
coronal fields. To further explore this possibility,
we plan to simulate this patchy reconnection in a two-dimensional
Y-type current sheet. In this case, we expect that the reconnected tubes
will decelerate as they descend toward the Y-type null and the arcade
of field lines below it, much as the coronal voids are observed to do
\citep{Sheeley2004}.  We will compare the deceleration 
of the simulated tubes with the observed deceleration of the
coronal voids, and therefore
more accurately test the validity of this model.
We will also simulate multiple, simultaneous reconnection
events in the same current sheet studying how these
tubes interact with each other, and whether their final equilibrium
resembles the tangled, sheared equilibrium formed by post-flare
coronal loop arcades.

\acknowledgements

We wish to thank Drs. J. Klimchuk, D. McKenzie, N. Sheeley, 
and H. Warren
for useful discussions.  This work was carried out with support 
from NASA and ONR, and was in part completed while the authors
were participating in programs at the Kavli Institute for 
Theoretical Physics and at the Newton Institute.

\eject

\eject

\begin{figure}[ht]
\epsscale{1.}
\plottwo{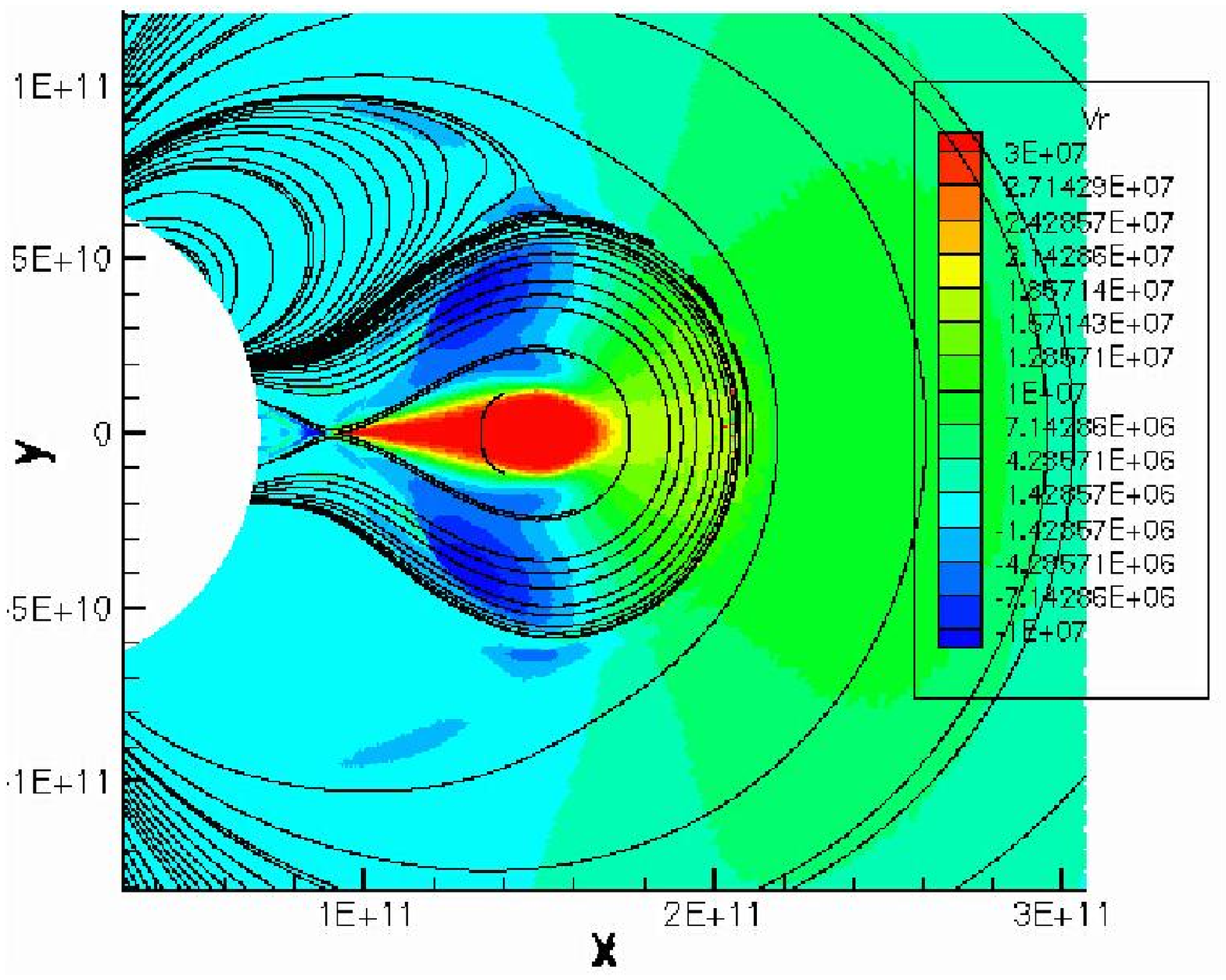}{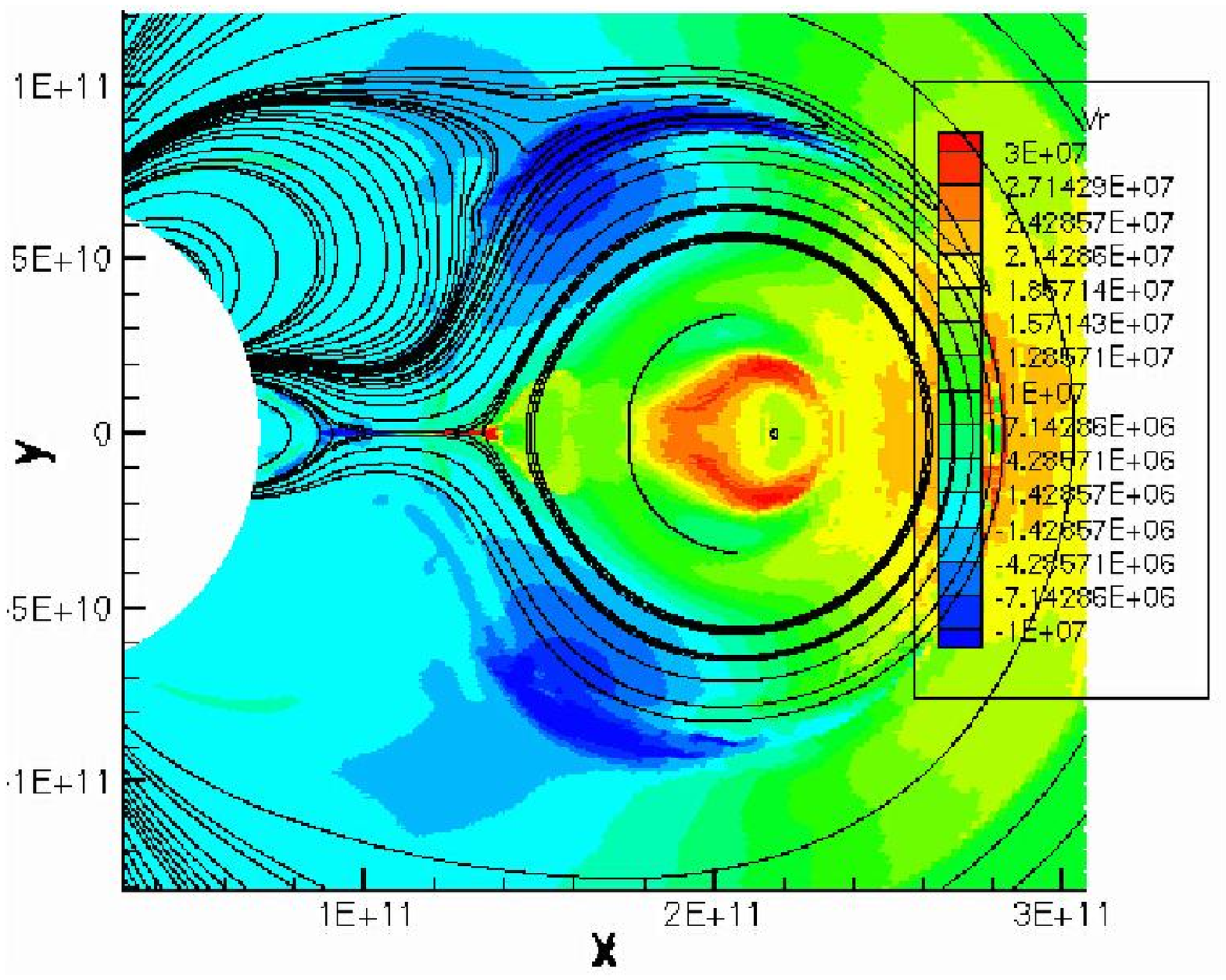}
\caption{Snapshots from a two dimensional MHD simulation of a
coronal mass ejection, showing the geometry of the fields during 
eruption. The field bows out from the
low corona in the left panel, erupting from the solar atmosphere. Once
the eruption is fully developed, the field starts to pinch off
and reconnect on the back side of the CME, as shown
in the right hand panel. The reconnected
fields on the right hand side of this newly formed current
sheet spring up to the CME, while those on the left side
spring down into the corona to form arcade loops.
From \citet{MacNeice2004}.
\label{fig:cme}}
\end{figure}

\begin{figure}[ht]
\epsscale{2.}
\includegraphics[width=4in,angle=90]{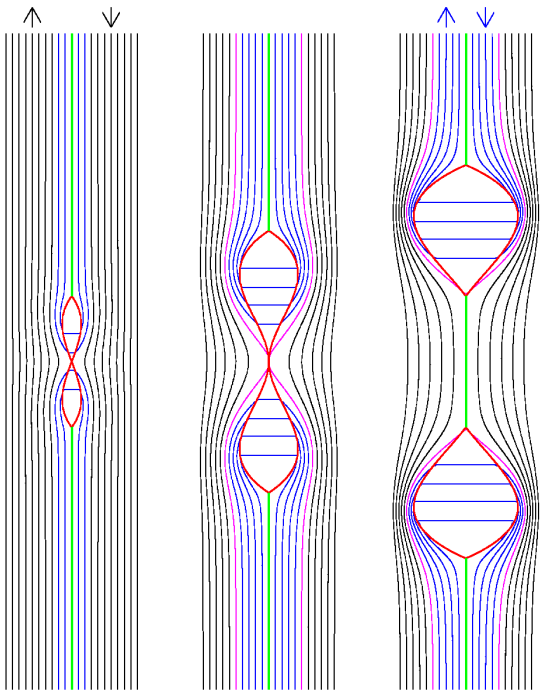}
\caption{Theoretical field configuration for a burst of reconnection 
in 2D.  The red lines show the slow shock fronts, the green lines
show the current sheet, the blue lines are reconnected field lines,
the black lines are unreconnected field lines, and the pink
lines are the separatrices between the two.
From \citet{Longcope2004},  after \citet{Biernat1987}. 
\label{fig:biernat}}
\end{figure}

\begin{figure}[ht]
\epsscale{1.}
\plotone{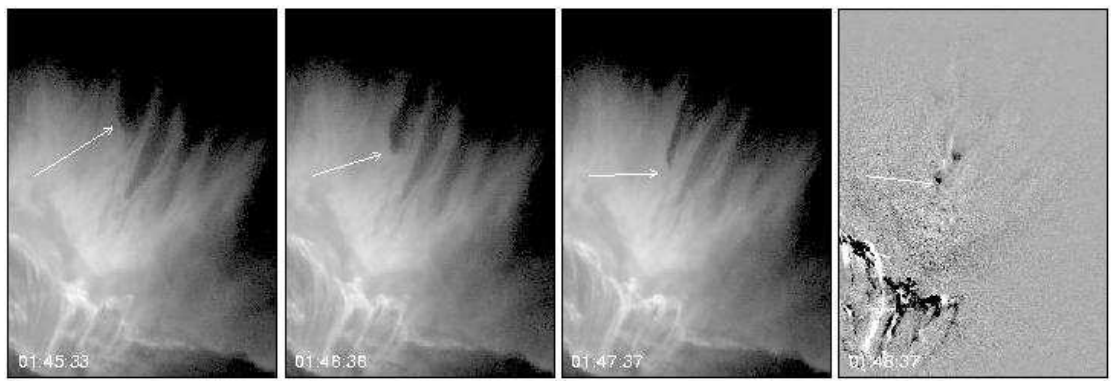}
\caption{
TRACE Images from the April 21, 2002 X flare. The first three
panels show unsubtracted images from the flare. The diffuse, bright
emission corresponds to high temperature plasma from the
Fe XXIV 192 \AA\ line in the TRACE 195 \AA\ bandpass. The
loop-like structures on the bottom of the arcade are loops that have
cooled down to about 1 MK and are emitting in Fe XII  195 \AA.
The arrow points to a void that is descending through the
hot plasma cloud. The final panel is a difference image between
successive TRACE exposures. From \citet{Sheeley2004}.
\label{fig:sheeley}}
\end{figure}

\begin{figure}[ht]
\epsscale{.75}
\includegraphics[width=4in,angle=90]{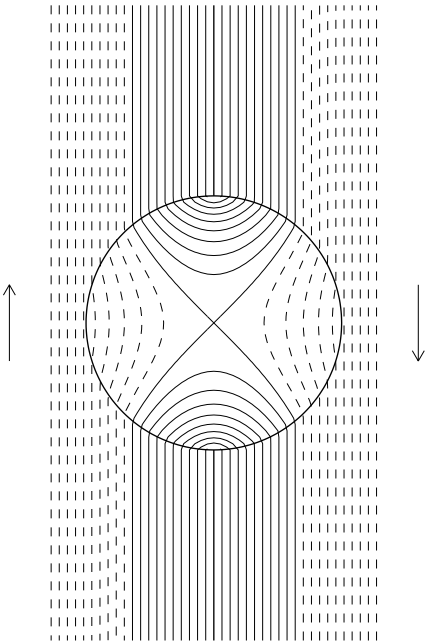}
\caption{Contours of the normalized flux function $A_0(x,z)$,
from the real part of Equation \ref{eq:2dpotential}, showing
field lines after a two dimensional reconnection episode.  
The reconnection region is enclosed by a thick sold line, 
inside which the field is current-free.  Solid lines show the 
field lines in the two reconnected flux tubes; dashed lines show 
the unreconnected field lines.}
\label{fig:2d}
\end{figure}

\begin{figure}[htp]
\epsscale{1.0}
\plotone{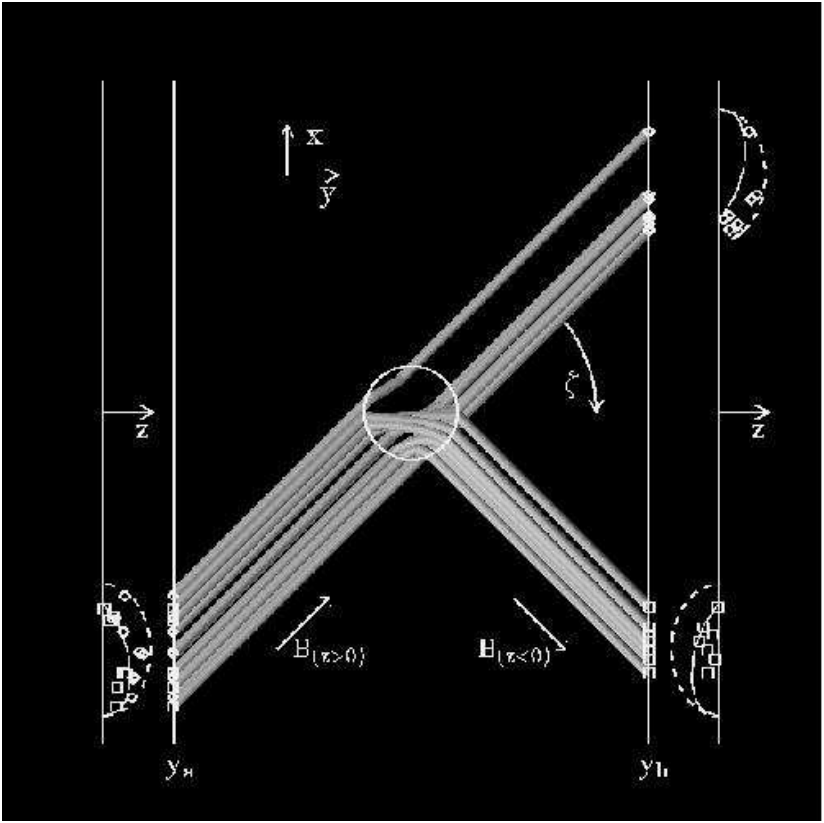}
\caption{3D plot of field lines intersecting 
the sphere of potential reconnected field. The field lines 
shown all cross the $y_a<0$ plane above the current sheet 
($z>0$). When the field lines pass through the reconnection
sphere they deviate from their straight trajectory. Some are
unreconnected, and so cross to $y_b>0$
plane above the current sheet, but others 
have reconnected and so cross the $y_b>0$ plane below
the current sheet. The points at which
they intersect the two $y=constant$ planes are shown on 
the left and right sides as $x-z$ projections, along
with the separatrix curve which separates reconnected
from unreconnected field lines in those planes (solid lines),
and a dashed curve delineating the edge of the flux which
intersects the reconnection sphere.}
    \label{fig:3dfl}
\end{figure}

\begin{figure}[htp]
\epsscale{1.}
\plotone{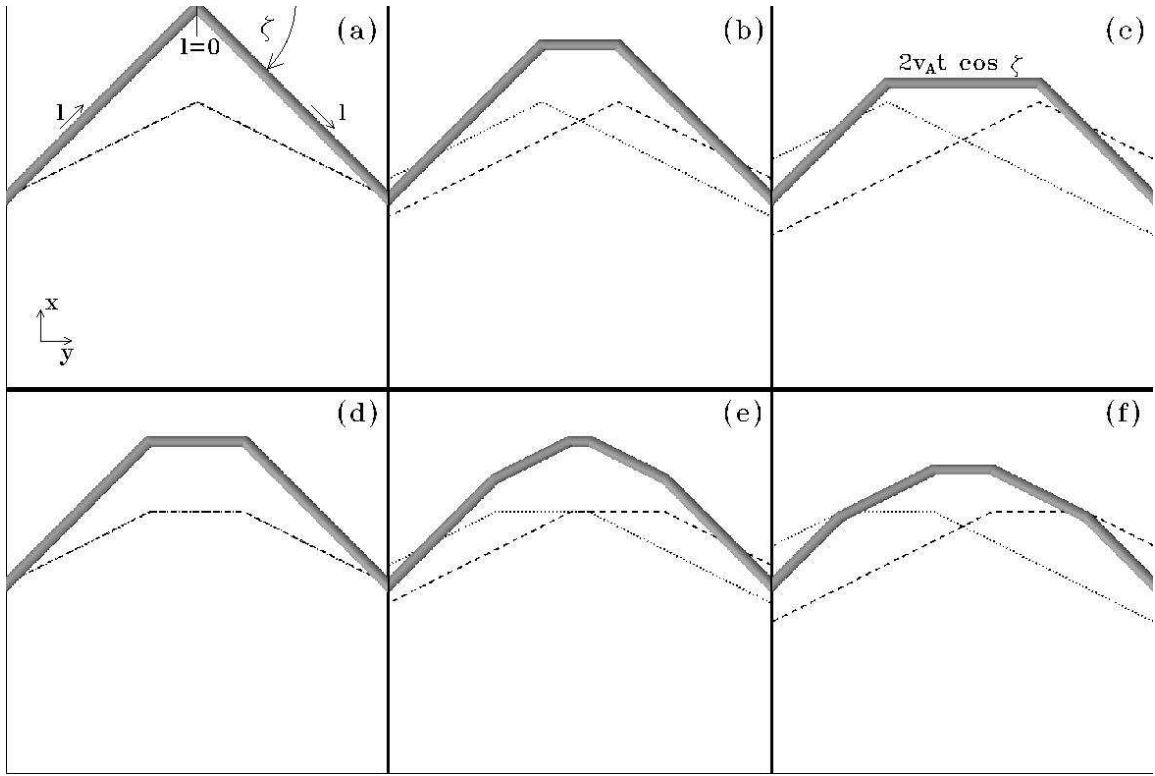}
\caption{The evolution of two post-reconnection flux tubes
according to the thin flux tube dynamics of \S 3.1.  A central 
segment, between the slow mode shocks formed at
the bends in the tube, moves vertically down due to 
magnetic tension, while the rest of the tube beyond
the slow shocks remains fixed.  Panels {\it a}-{\it c} show the motion 
for a flux tube initially forming the upper part of a triangle, 
while panels {\it d}-{\it f} show the motion for a flux tube initially 
forming the upper part of a trapezoid.}
    \label{fig:tft}
\end{figure}

\begin{figure}[ht]
\epsscale{1.}
\plotone{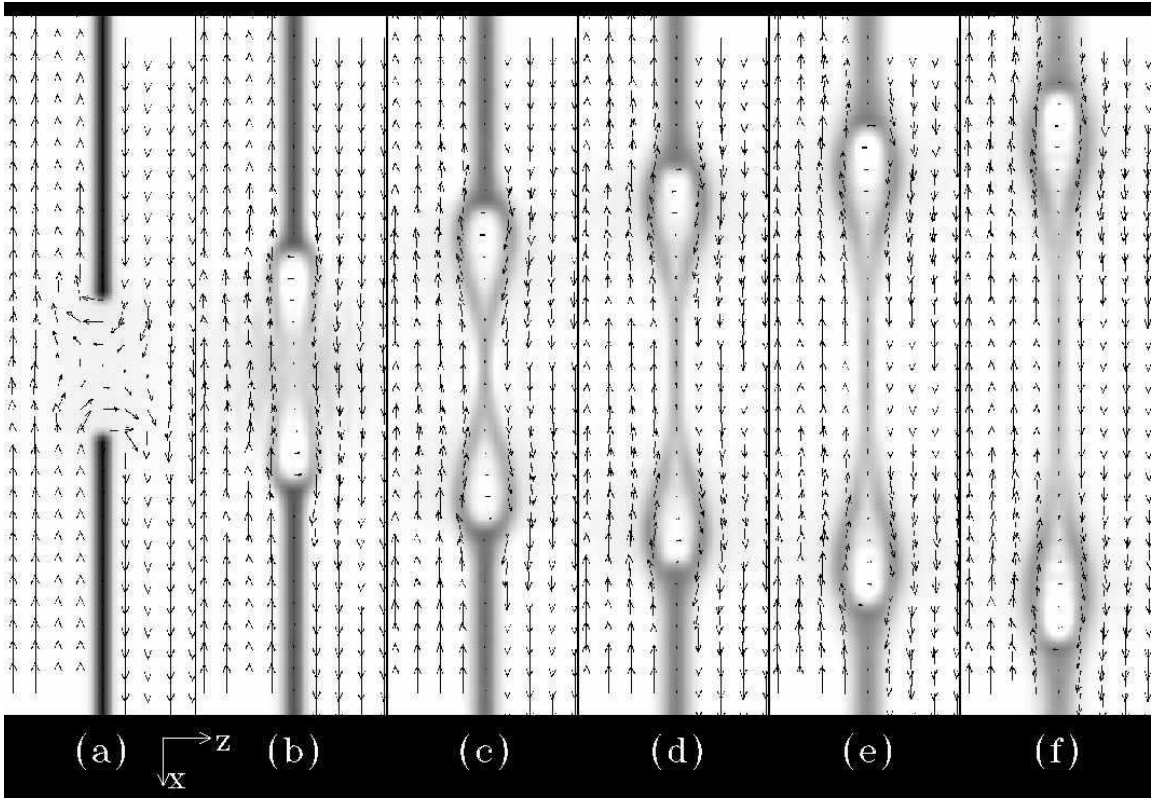}
\caption{Slices at $y=0$ of magnetic field from the potential 
reconnection simulation at $\zeta=2\pi/8$, at times
$tv_{A\perp}/L=[0.0, 0.2, 0.4, 0.6, 0.8, 1.0]$. 
The vectors show the in-plane magnetic field, while the
greyscale shows the guide field ($B_y$), where black
is zero field and white is positive.  Panel {\it a} shows
the initial configuration, including the potential sphere at the
center. Panels {\it b}-{\it f} show the subsequent evolution, as
the reconnected fields pull away from this sphere and
form into teardrop shapes.
\label{fig:bbla_b}}
\end{figure}

\begin{figure}[ht]
\epsscale{1.}
\plotone{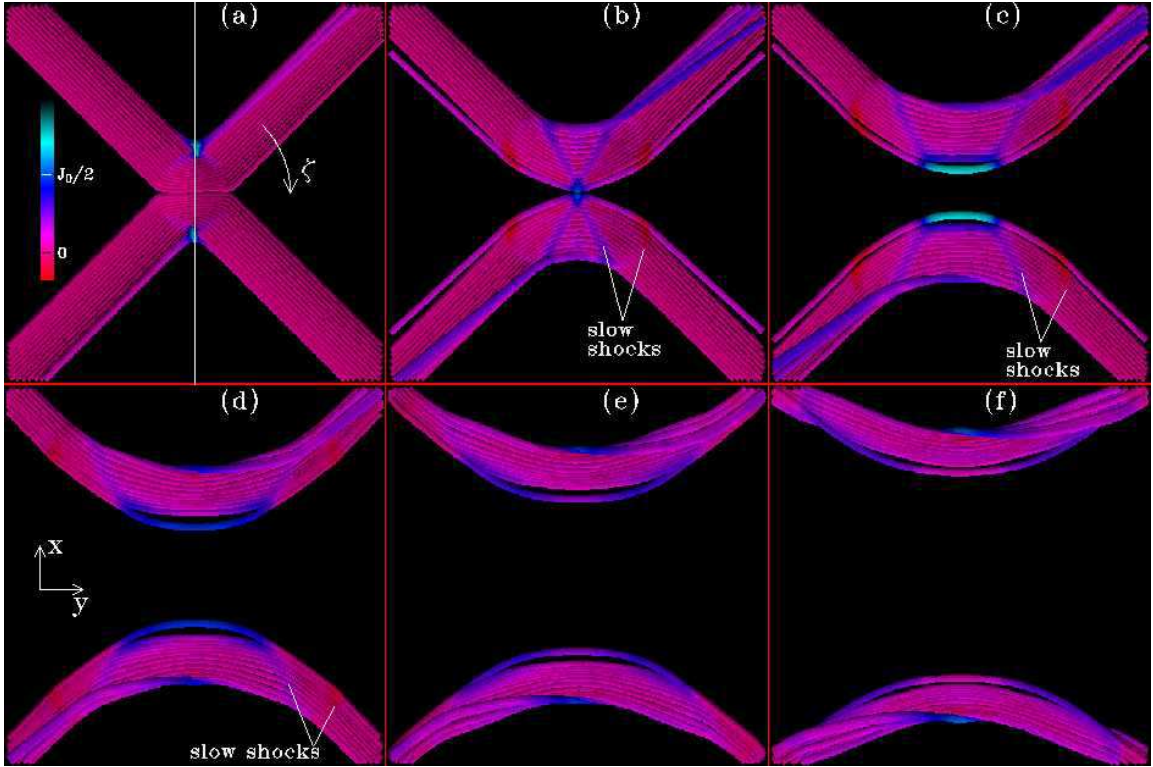}
\caption{Three dimensional view of the magnetic field lines
from the potential reconnection simulation at $\zeta=2\pi/8$,
at the same times as in Figure \ref{fig:bbla_b}. 
Field lines are traced from Lagrangian particles initially
on the $y=0$ plane inside the reconnection sphere. The color
is set by the level of current density parallel to the
magnetic field, normalized by the maximum initial current
density $J_0$. The slow mode shocks set up by the reconnection
are visible in panels {\it b}-{\it d}. The white vertical line in
panel {\it a} shows the location of the $y=0$ cuts of Figure
\ref{fig:bbla_b}.
\label{fig:bbla_fl}}
\end{figure}

\begin{figure}[ht]
\epsscale{1.}
\plotone{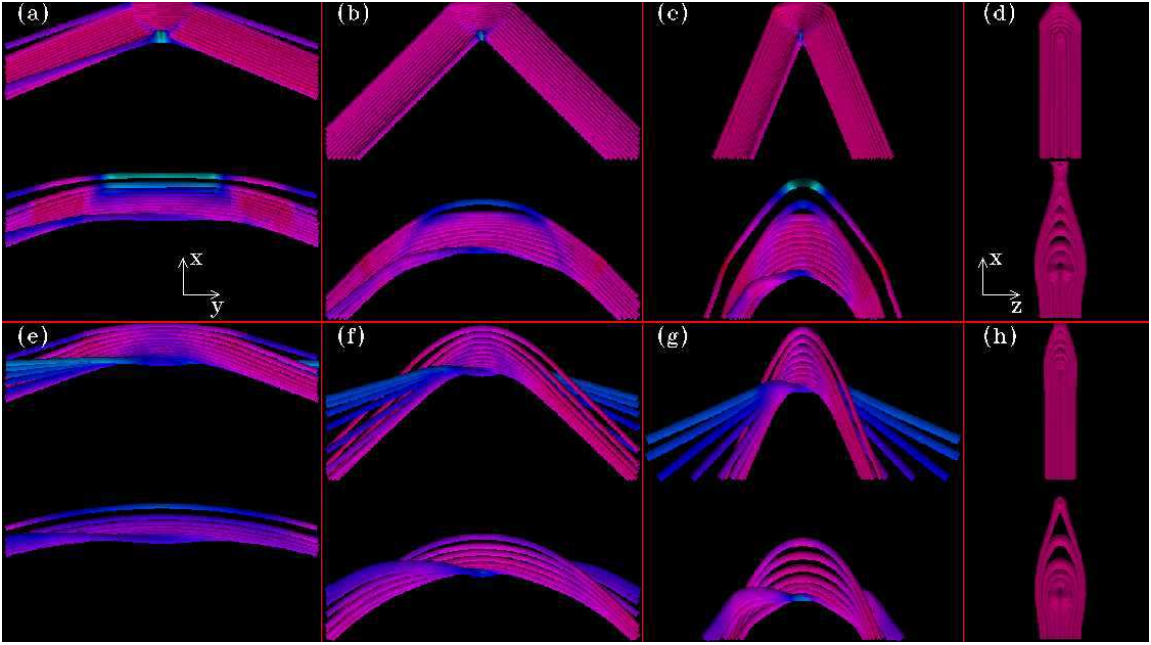}
\caption{Field lines of the high-$\beta$ 
reconnection simulations at the four angles simulated:
$\zeta=[1,2,3,4]\pi/8$, from left to right.
The top halves of panels {\it a}-{\it d} show the potential simulations 
in their initial state, while the bottom halves of these 
panels show the tubes during their evolution at 
$tv_{A\perp}/L=[0.3,0.6,0.8,0.8]$, respectively.
The top halves of panels {\it e}-{\it g} show the resistive simulations 
after the same amount of flux has reconnected as was initially 
reconnected in the potential simulations, at $tv_{A\perp}/L=[0.2,0.3,0.5]$,
respectively. Because the flux at 
$\zeta=4\pi/8$ evolves so quickly once reconnected, the top half 
of panel (h) is shown at an earlier time, $tv_{A\perp}/L=0.3$, when the 
reconnected flux equals only half of that reconnected in the potential state. 
The bottom halves of panels {\it e}-{\it h} show the reconnected
tubes later in their evolution, at $tv_{A\perp}/L=[0.6,0.9,1.1,0.8]$, 
respectively.
Note that panels {\it a}-{\it c} and {\it e}-{\it g}
show the field lines from the $-\zhat$ direction
while panels {\it d} and {\it h}, where the guide field is zero, 
show the field lines from the $\yhat$ direction.
\label{fig:bblsumm}}
\end{figure}

\begin{figure}[ht]
\epsscale{.75}
\plotone{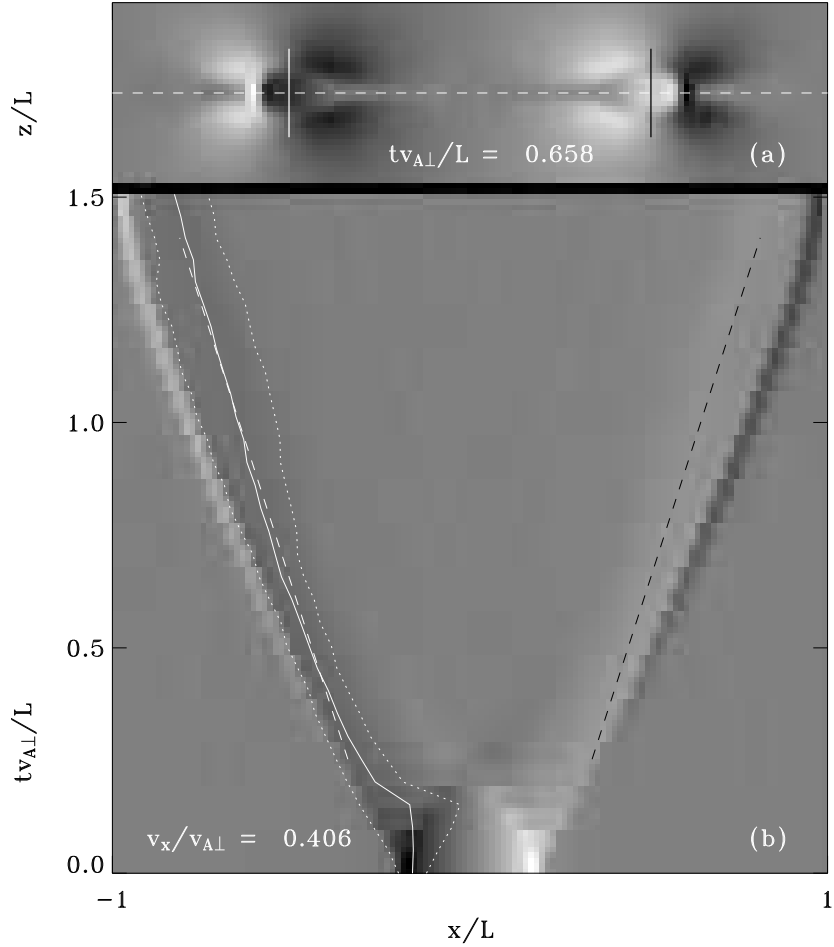}
\caption{Height versus time stackplot. Panel {\it a} shows a $y=0$ 
slice of $B_z$ during the potential reconnection
simulation at $\zeta=2\pi/8$. A gaussian
is fit to the negative (black) part of $B_z$ along the white
dashed curve for $x<0$, and this is used to find the center
of mass (shown as the white vertical line) of the teardrop shape.
Panel {\it b} shows cuts along this same dashed line at
successive times, stacked on top of each other to give
a time-distance plot. The time-distance location of
the center of mass is shown as the solid line in this
panel, while the $\pm 2\sigma$ level of the gaussian
fit is shown as the two dotted lines. The linear least
squares fit to the center of mass location, taken
from the time the teardrop shape is first fully formed
until it first reaches the simulation boundary, is shown
as the dashed line. The slope of this line gives the
velocity of the reconnected flux tube, $v=0.406v_{A\perp}$.
\label{fig:ht}}
\end{figure}

\begin{figure}[ht]
\epsscale{1.}
\plotone{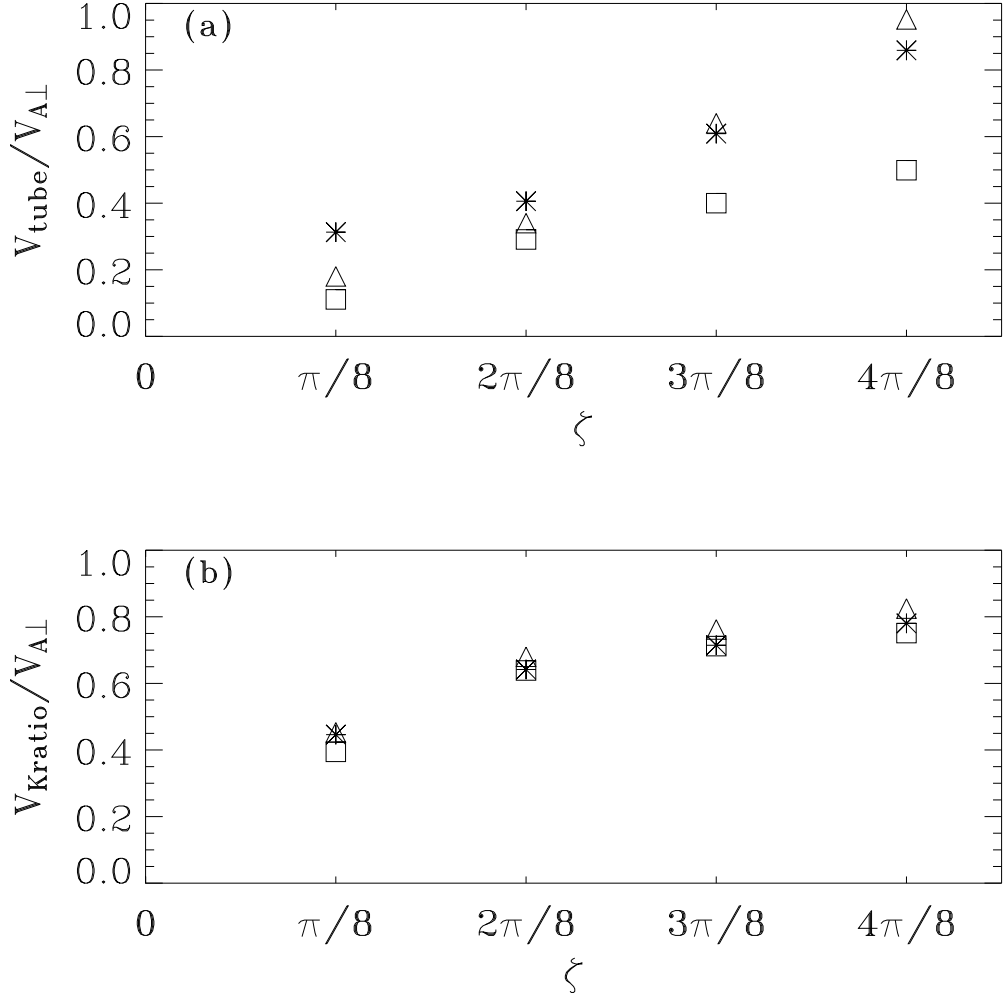}
\caption{Panel {\it a}: measured velocities of the reconnected flux
tube velocities. Panel {\it b}: predictions of this velocity based 
on the added mass calculation of \S 3.3. The asterisks 
show the velocities from the high $\beta$ potential reconnection 
simulations with a dip in the guide field at the current sheet. 
The open triangles show the velocities from the high-$\beta$ 
resistive reconnection simulations with the same dipped 
guide field. The open squares show the velocities from
the low-$\beta$ resistive reconnection simulations. 
\label{fig:velocity}}
\end{figure}

\begin{figure}[ht]
\epsscale{1.}
\plotone{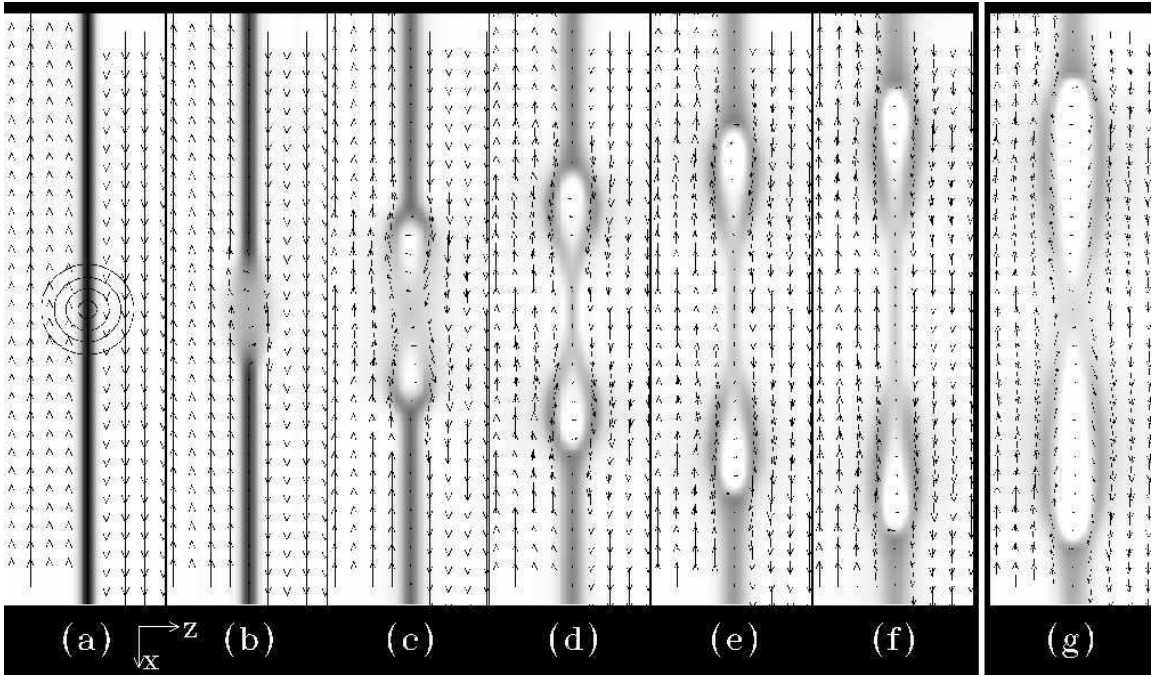}
\caption{ Two dimensional magnetic field slices for
the high-$\beta$ resistive simulation at $\zeta=2\pi/8$,
in the format of Figure \ref{fig:bbla_b}.
Panels {\it a} through {\it g} show slices from this
simulation at $tv_{A\perp}/L=[0.0,0.05,0.34,0.65,0.96,1.27]$.
This shows the evolution of magnetic field which reconnects in a
small, 3D sphere in the center of a 1D current sheet. 
The contours of resistivity are shown in panel {\it a} at 
$10$, $30$, $60$, and $90$ times the background level of 
$\eta_0/(\delta v_A)=1/200$. 
The resistive sphere is turned off after panel {\it c}, 
as can be seen by the reformation of the thin current sheet at the
center in panel {\it d}.
The 3D view of this simulation is shown in Figure 
\ref{fig:neta2f_fl}.
Panel {\it g} shows a cut at $tv_{A\perp}/L=1.29$
from a different simulation with the same 
initial conditions, but in which the resistive sphere is 
never turned off.
\label{fig:neta2f_b}}
\end{figure}

\begin{figure}[ht]
\epsscale{1.}
\plotone{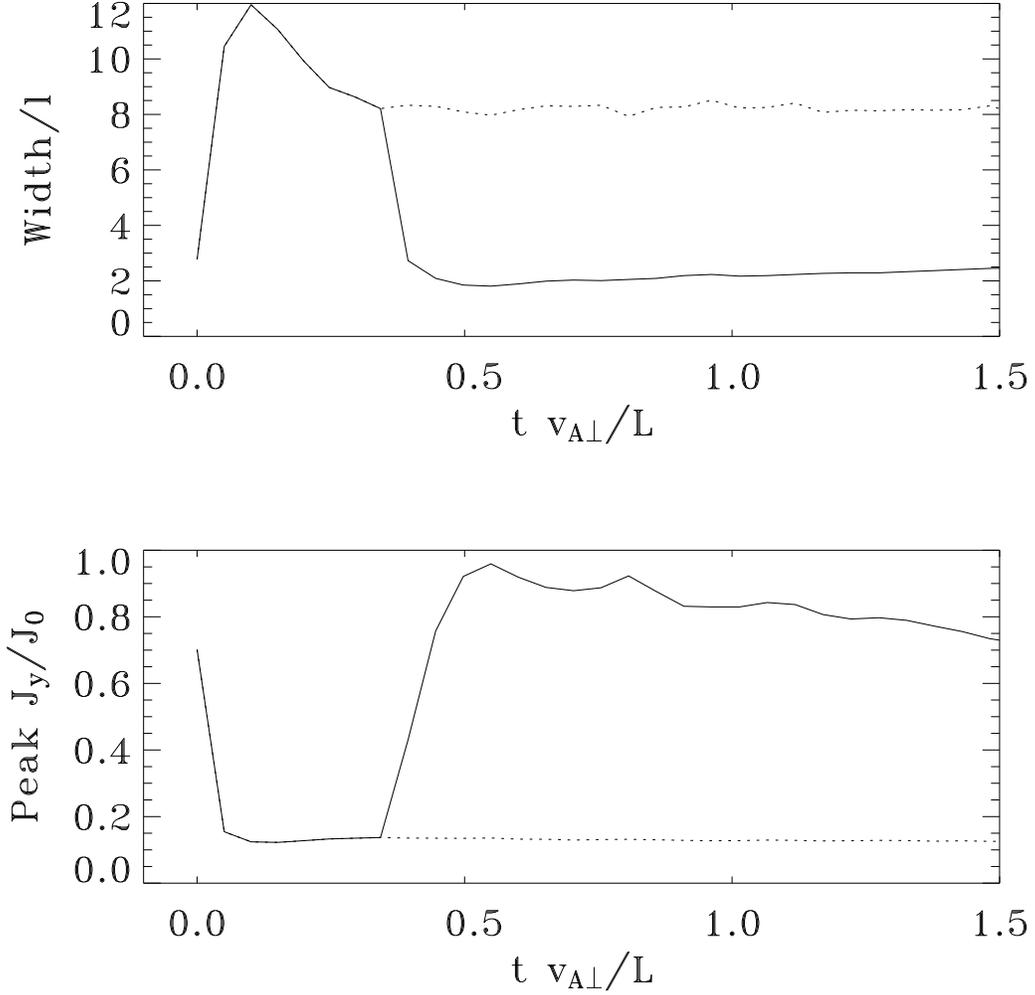}
\caption{ Time variation of the current sheet along
the $x=y=0$ line through the center of the high-$\beta$ 
resistive reconnection simulation of Figures
\ref{fig:neta2f_b} and \ref{fig:neta2f_fl}.
Panel {\it a} shows the full width at half of the maximum
current value as the solid line.  Panel {\it b} shows
peak current strength in the ${\bf{\hat y}}$ direction (solid). 
The dotted lines in both panels show the width and amplitude 
of the current sheet if the resistivity in the sphere is 
never turned off.
\label{fig:sheet}}
\end{figure}

\begin{figure}[ht]
\epsscale{.90}
\plotone{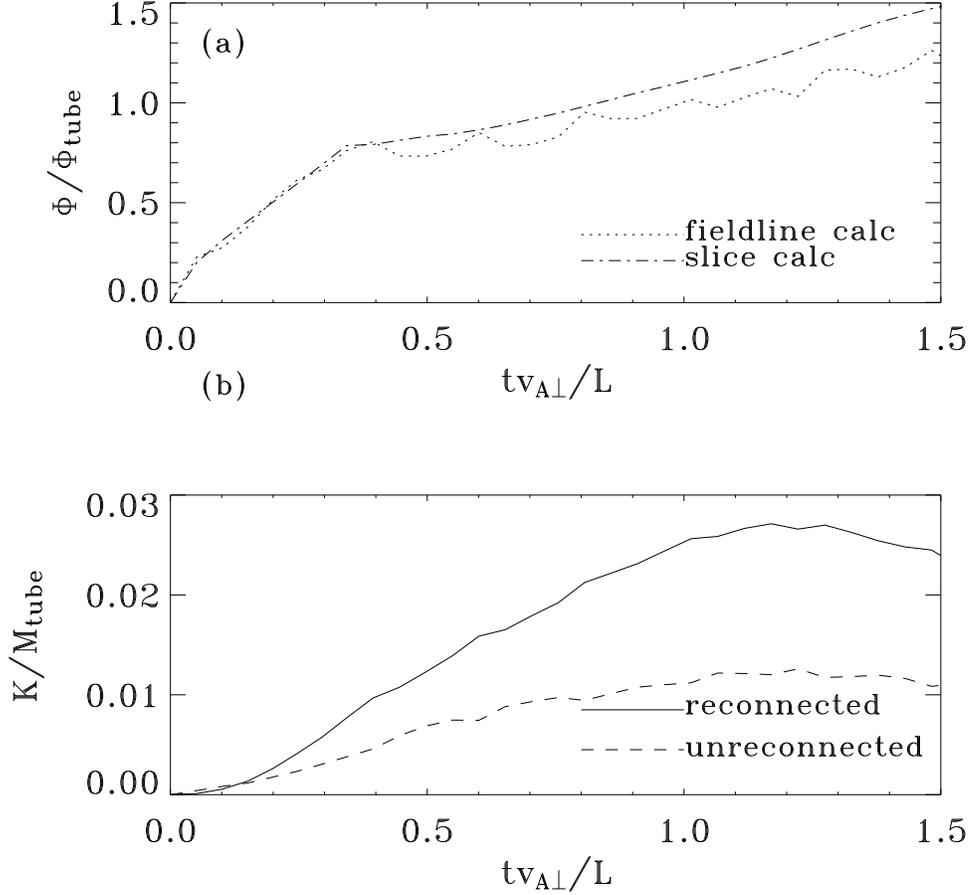}
\caption{
Panel {\it a} shows the measured reconnected flux in the high-$\beta$
resistive reconnection simulation at $\zeta=2\pi/8$, normalized
by $\Phi_{tube}$ from Equation \ref{eq:phi_tube}.
The dash-dotted line shows the flux calculated from summing
over negative $B_z$ on the $z=0$ plane at $x<0$.
The dotted line shows the flux calculated from the field line 
summing routine of \S 5.  Panel {\it b} shows
the global kinetic energies of reconnected (solid) versus unreconnected 
(dashed) field lines, taken from the fieldline summing routine.
These are normalized by the magnetic energy of the two unreconnected
flux tubes initially intersecting the reconnection sphere.
\label{fig:globals}}
\end{figure}

\begin{figure}[ht]
\epsscale{1.}
\plotone{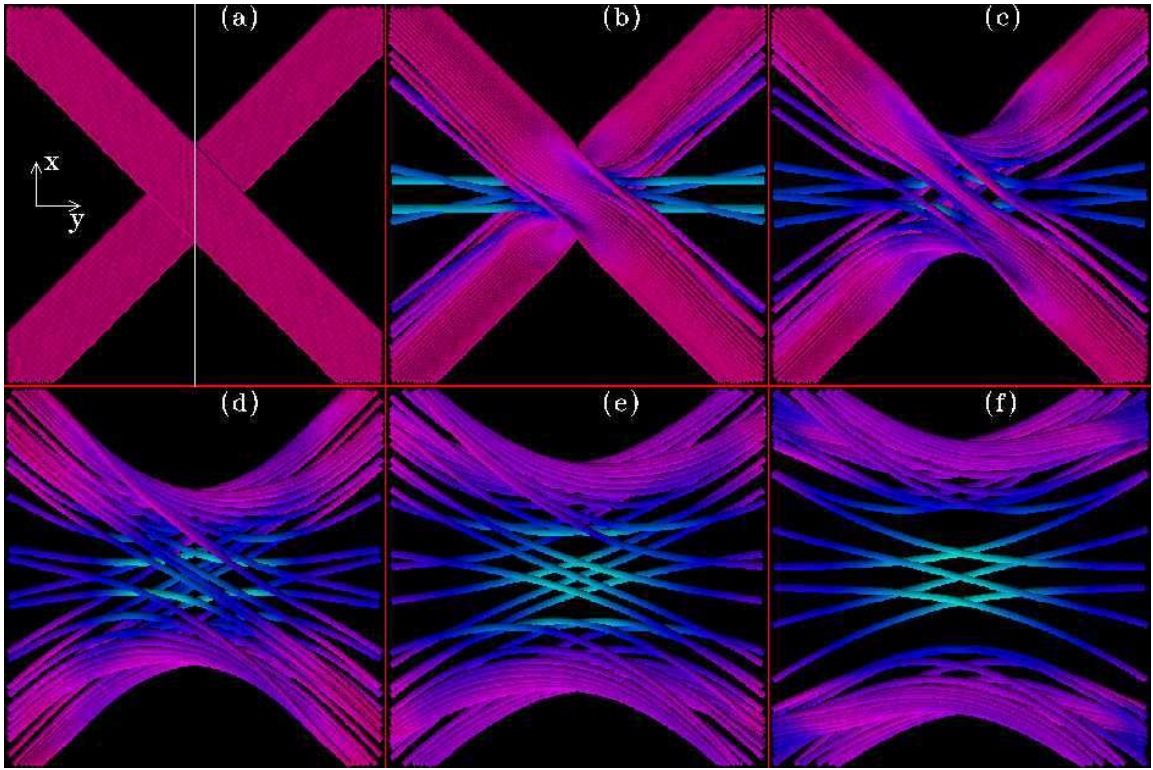}
\caption{ Three dimensional view of field lines from the
high-$\beta$ resistive reconnection simulation at $\zeta=\pi/4$,
at times $tv_{A\perp}/L=[ 0.00,0.30,0.60,0.91,1.22,1.53]$.
This shows a representative set of field lines which reconnect in the 
small sphere of high resistivity on the initially 1D current
sheet to form two arched loops. The white line in panel {\it a} shows
the location of the 2D plane of Figure \ref{fig:neta2f_b}.
\label{fig:neta2f_fl}}
\end{figure}

\begin{figure}[ht]
\epsscale{1.}
\plotone{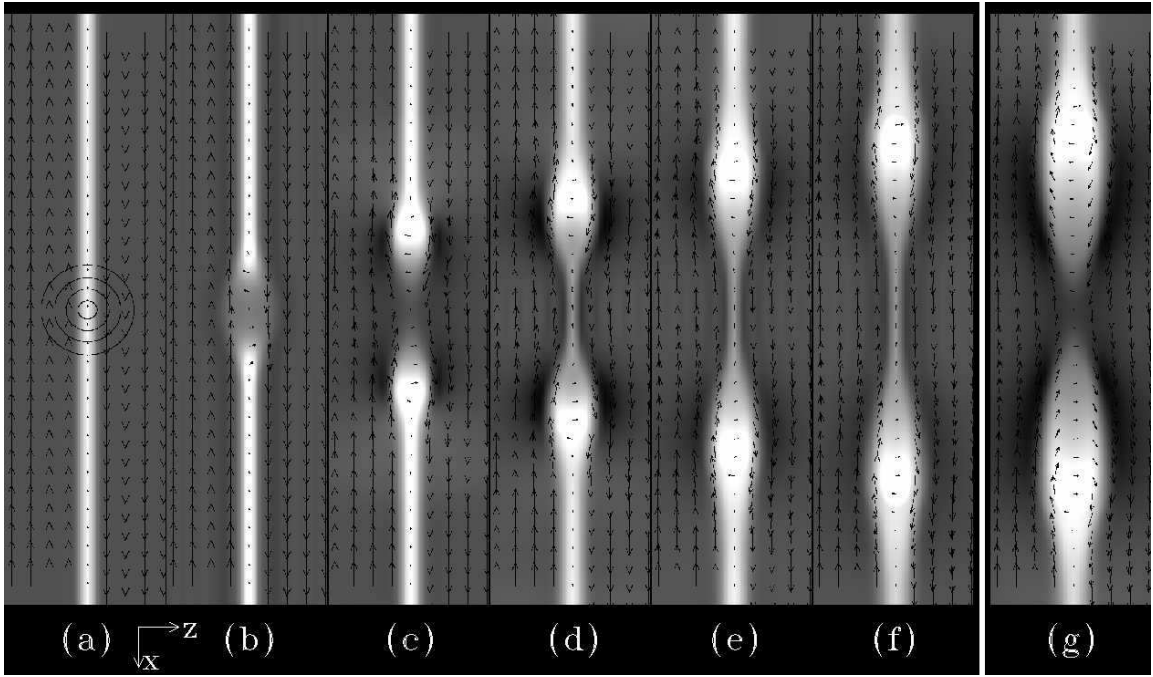}
\caption{ Two dimensional view of the magnetic field in the 
low-$\beta$ resistive reconnection simulation at $\zeta=\pi/4$,
in the format of Figure \ref{fig:bbla_b}. Panels {\it a} through
{\it f} show this simulation at times 
$tv_{A\perp}/L=[ 0.00,0.06,0.34,0.62,0.91,1.20]$.
In this case,
the current sheet appears white because the guide field
is strongest there (to balance the drop in the reconnection
component of the field). This contrasts with Figure \ref{fig:bbla_b}
and \ref{fig:neta2f_b}, where the guide field drops to zero
in the current sheet. The field evolves much as it does
in the equivalent high-$\beta$ resistive reconnection
simulation shown in Figure \ref{fig:neta2f_b}. Again,
the resistive sphere is turned off after panel {\it c},
and panel {\it g} shows a second simulation,
in which the resistivity is never turned off, at
$tv_{A\perp}/L=1.20$.
\label{fig:lbeta2d_b}}
\end{figure}


\begin{thebibliography}{}

\bibitem[{Asai} et~al., 2004]{Asai2004}
{Asai}, A., {Yokoyama}, T., {Shimojo}, M., \& {Shibata}, K. 2004,
\newblock \apjl, 605, L77

\bibitem[Biernat et~al., 1987]{Biernat1987}
Biernat, H.~K., Heyn, M.~F., \& Semenov, V.~S. 1987,
\newblock JGR, 92, 3392

\bibitem[{Birn} et~al., 2001]{Birn2002}
{Birn}, J. et~al. 2001,
\newblock JGR, , 3715

\bibitem[Biskamp, 1986]{Biskamp1986}
Biskamp, D. 1986,
\newblock Phys. Fluids, 29, 1520

\bibitem[Biskamp \& Schwarz, 2001]{Biskamp2001}
Biskamp, D. \& Schwarz, E. 2001,
\newblock Phys. Plasmas, 8, 4729

\bibitem[Carmichael, 1964]{Carmichael1964}
Carmichael, H. 1964,
\newblock in AAS-NASA Symposium on the Physics of Solar Flares,  ed. W.~N.
  Hess, (NASA SP-50),  451

\bibitem[Dahlburg \& Norton, 1995]{Dahlburg1995}
Dahlburg, R.~B. \& Norton, D. 1995,
\newblock in Small Scale Structures in Three-Dimensional Hydrodynamic and
  Magnetohydrodynamic Turbulence,  ed. M.~Meneguzzi, A.~Pouquet, \& P.~Sulem,
  (Heidelberg: Springer-Verlag),  331

\bibitem[Erkaev et~al., 2000]{Erkaev2000}
Erkaev, N.~V., Semenov, V.~S., \& Jamitsky, F. 2000,
\newblock Phys. Rev. Let., 84, 1455

\bibitem[Forbes, 2000]{Forbes2000}
Forbes, T. 2000,
\newblock JGR, 105, 23153

\bibitem[{Forbes} \& {Acton}, 1996]{Forbes1996}
{Forbes}, T.~G. \& {Acton}, L.~W. 1996,
\newblock \apj, 459, 330

\bibitem[{Gallagher} et~al., 2002]{Gallagher2002}
{Gallagher}, P.~T., {Dennis}, B.~R., {Krucker}, S., {Schwartz}, R.~A., \&
  {Tolbert}, A.~K. 2002,
\newblock \solphys, 210, 341

\bibitem[Heyn \& Semenov, 1996]{Heyn1996}
Heyn, M. \& Semenov, V. 1996,
\newblock Phys. Plasmas, 3, 2725

\bibitem[Hirayama, 1974]{Hirayama1974}
Hirayama, T. 1974,
\newblock Sol. Phys., 34, 323

\bibitem[Hirose et~al., 2001]{Hirose2001}
Hirose, S., Uchida, Y., Uemure, S., Yamaguchi, T., \& Cable, S.~B. 2001,
\newblock ApJ, 551, 586

\bibitem[Hornig, 2005]{HornigN2005}
Hornig, G. 2005,
\newblock in Reconnection of Magnetic Fields: Magnetohydrodynamics and
  Collisionless Theory and Observations,  ed. J.~Birn \& E.~R. Priest,
  (Cambridge, UK: Cambridge University Press), pages 24, in press

\bibitem[Huba, 2005]{Huba2005}
Huba, J.~D. 2005,
\newblock Phys. Plasmas, 12, 12322

\bibitem[{Innes} et~al., 2003]{Innes2003a}
{Innes}, D.~E., {McKenzie}, D.~E., \& {Wang}, T. 2003,
\newblock \solphys, 217, 267

\bibitem[Klimchuk, 2001]{Klimchuk2001}
Klimchuk, J.~A. 2001,
\newblock in Space Weather,  ed. .~H.~S. P.~Song, G.~Siscoe, volume Geophysical
  Monograph 125, (Washington: AGU),  142

\bibitem[Kopp \& Pneuman, 1976]{Kopp1976}
Kopp, R.~A. \& Pneuman, G.~W. 1976,
\newblock Sol. Phys., 50, 85

\bibitem[Kulsrud, 2001]{Kulsrud2001}
Kulsrud, R.~M. 2001,
\newblock Earth, Planets and Space, 53, 417

\bibitem[Lin et~al., 2003]{Lin2003}
Lin, J., Soon, W., \& Baluinas, S. 2003,
\newblock JGR, 105, 2375

\bibitem[Linton \& Antiochos, 2005]{LintonA2005}
Linton, M.~G. \& Antiochos, S.~K. 2005,
\newblock ApJ, 625, 506

\bibitem[Longcope, 2004]{Longcope2004}
Longcope, D.~W. 2004,
\newblock Quantifying magnetic reconnection and the heat it generates,
\newblock in Proceedings of the SOHO 15 Workshop -- Coronal Heating,  ed. R.~W.
  Walsh, J.~Ireland, D.~Danesy, \& B.~Fleck, volume 575 of {\em ESA SP}, pages
  198--209, Paris, European Space Agency

\bibitem[Low, 2001]{Low2001}
Low, B.~C. 2001,
\newblock JGR, 106, 25141

\bibitem[MacNeice et~al., 2004]{MacNeice2004}
MacNeice, P., Antiochos, S.~K., Phillips, A., Spicer, D.~S., DeVore, C.~R., \&
  Olson, K. 2004,
\newblock ApJ, 614, 1028

\bibitem[{McKenzie} \& {Hudson}, 1999]{McKenzie1999}
{McKenzie}, D.~E. \& {Hudson}, H.~S. 1999,
\newblock \apjl, 519, L93

\bibitem[Nitta et~al., 2002]{Nitta2002}
Nitta, S., Tanuma, S., \& Maezawa, K. 2002,
\newblock ApJ, 580, 538

\bibitem[Nitta et~al., 2001]{Nitta2001}
Nitta, S., Tanuma, S., Shibata, K., \& Maezawa, K. 2001,
\newblock ApJ, 550, 1119

\bibitem[Parker, 1957]{Parker1957}
Parker, E.~N. 1957,
\newblock JGR, 62, 509

\bibitem[Petschek, 1964]{Petschek1964}
Petschek, H.~E. 1964,
\newblock , NASA-SP-50, 425

\bibitem[Pontin et~al., 2005]{Pontin2005}
Pontin, D.~I., Galsgaard, K., Hornig, G., \& Priest, E.~R. 2005,
\newblock Ph. Plasmas, 12, 052307

\bibitem[Pontin et~al., 2004]{Pontin2004}
Pontin, D.~I., Hornig, G., \& Priest, E.~R. 2004,
\newblock Geoph. Atroph. Fluid Dynamics, 98, 407

\bibitem[Scholer \& Roth, 1987]{Scholer1987}
Scholer, M. \& Roth, D. 1987,
\newblock JGR, 92, 3223

\bibitem[Semenov et~al., 1983]{Semenov1983}
Semenov, V.~S., Heyn, M.~F., \& Kubyshkin, I.~V. 1983,
\newblock Sov.\ Astron., 27, 600

\bibitem[Semenov et~al., 1998]{Semenov1998}
Semenov, V.~S., Volkonskaya, N.~N., \& Biernat, H.~K. 1998,
\newblock Phys. Plasmas, 5, 3242

\bibitem[{Sheeley} et~al., 2004]{Sheeley2004}
{Sheeley}, N.~R., {Warren}, H.~P., \& {Wang}, Y.-M. 2004,
\newblock \apj, 616, 1224

\bibitem[Spruit, 1981]{Spruit1981}
Spruit, H.~C. 1981,
\newblock A\&A, 98, 155

\bibitem[Sturrock, 1966]{Sturrock1966}
Sturrock, P.~A. 1966,
\newblock Nature, 211, 695

\bibitem[Sweet, 1958]{Sweet1958}
Sweet, P.~A. 1958,
\newblock Nuovo Cimento Suppl., 8, 188

\bibitem[Ugai \& Tsuda, 1977]{Ugai1977}
Ugai, M. \& Tsuda, T. 1977,
\newblock J. Plasma Phys., 17, 337

\end{thebibliography}
\end{document}